\documentclass[12pt]{article}
\usepackage{graphicx}
\usepackage{enumerate}
\usepackage[sort]{natbib}
\usepackage{url} % not crucial - just used below for the URL 
\usepackage{header}   % our style file

\newcommand{\blind}{1}

\begin{document}

\def\spacingset#1{\renewcommand{\baselinestretch}%
{#1}\small\normalsize} \spacingset{1}

%%%%%%%%%%%%%%%%%%%%%%%%%%%%%%%%%%%%%%%%%%%%%%%%%%%%%%%%%%%%%%%%%%%%%%%%%%%%%%

\if1\blind
{
  \title{\bf A User-Friendly Computational Framework for Robust Structured Regression with the L$_2$ Criterion}
  \date{}
  \author{Jocelyn T. Chi
%%  	\hspace{.2cm}\\
%%    Department of Mathematics, University of California at Los Angeles\\
%%    and \\
	and 
    Eric C. Chi\thanks{
    	The work was supported in part by NSF grants DMS-1760374, DMS-1752692, and DMS-2103093.}
%\\
%    Department of Statistics, Rice University}
	}
  \maketitle
} \fi

\if0\blind
{
  \bigskip
  \bigskip
  \bigskip
  %\begin{center}
    \title{\bf A User-Friendly Computational Framework for Robust Structured Regression Using the L$_2$ Criterion}
    \date{}
    \author{}
	%\end{center}
	\maketitle
  \medskip
} \fi

\bigskip
\begin{abstract}
We introduce a user-friendly computational framework for implementing robust versions of a wide variety of structured regression methods with the L$_{2}$ criterion.  In addition to introducing an algorithm for performing L$_{2}$E regression, our framework enables robust regression with the L$_{2}$ criterion for additional structural constraints, works without requiring complex tuning procedures on the precision parameter, can be used to identify heterogeneous subpopulations, and can incorporate readily available non-robust structured regression solvers.  We provide convergence guarantees for the framework and demonstrate its flexibility with some examples.  Supplementary materials for this article are available online.
\end{abstract}

\noindent%
\vfill

\newpage
\spacingset{1.5} % DON'T change the spacing!

\section{Introduction}
\label{sec:intro}

Linear multiple regression is a classic method that is ubiquitous across numerous domains.  Its ability to accurately quantify a linear relationship between a response vector $\V{y} \in \Real$ and a set of predictor variables $\M{X} \in \Real^{n \times p}$, however, is diminished in the presence of outliers.  
The \LTE method \citep{Terrell1990, hjort1994, Sco2001, Sco2009} presents an approach to robust linear regression that optimizes the well-known L$_{2}$ criterion from nonparametric density estimation in lieu of the maximum likelihood.  Usage of the \LTE method for structured regression problems, however, has been limited by the lack of a simple computational framework.  
We introduce a general computational framework for performing a wide variety of robust structured regression methods with the L$_{2}$ criterion.  Our work offers the following novel contributions.
\begin{itemize}
	\item[1)] Our framework extends the \LTE method from \citet{Sco2001, Sco2009} to a wide variety of robust structured regression methods with the \LT criterion.
	\item[2)] Our framework enables simultaneous estimation of the regression coefficients and precision parameter as demonstrated in Section \ref{sec:computationalframework}.  We accomplish this via a block-coordinate descent algorithm.  Therefore, our simultaneous estimation simplifies the process of choosing a parameter that tunes the robustness of the estimation procedure. 
	%bypasses complex case-by-case procedures and additionally comes with algorithmic convergence guarantees.
	\item[3)] Our framework can be employed to ``robustify" existing implementations of non-robust structured regression methods in a ``plug-and-play" manner as detailed in Section \ref{sec:plugandplay} and demonstrated in Section \ref{sec:examples}.
	\item[4)] Our framework comes with convergence guarantees for the iterate sequence (Proposition \ref{prop:convergence}).  
\end{itemize}

We describe motivation for L$_{2}$ robust linear regression in Section \ref{sec:L2E}.  We introduce our computational framework with convergence guarantees in Section \ref{sec:computationalframework}.  We demonstrate the simplicity and flexibility of our framework by incorporating readily available structured regression solvers to implement robust versions of several MLE-based methods in Section \ref{sec:examples}.  Finally, we provide a brief discussion in Section \ref{sec:discussion}.

\subsection{Related Work}\label{sec:relatedwork}

The \LT minimization criterion has been used for histogram bandwidth selection as well as for obtaining kernel density estimators \citep{Sco1992}. Applying this well-known criterion from nonparametric density estimation to parametric estimation for regression problems enables a trade-off between efficiency and robustness in the estimation procedure.  In fact, \citet{BasHarHjo1998} introduced a family of divergences that includes the \LTE as a special case and the MLE as a limiting case. The members of this family of divergences are indexed by a parameter that explicitly trades off efficiency for robustness. While the MLE is the most efficient, it is also the least robust. Meanwhile, the \LTE represents a reasonable trade-off between efficiency and robustness \citep{warwick2005choosing}. The robustness of the \LTE can also be anticipated since it is a minimum distance estimator; estimators which are known to have robustness properties \citep{donoho1988automatic}.

Minimizing the \LT criterion has been used to develop robust statistical models including but not limited to quantile regression \citep{Lane2012}, mixture models \citep{Lee2010}, classification \citep{ChiScott2014}, forecast aggregation \citep{Ramos2014}, and survival analysis \citep{Yang2013}.  It has also found utility in engineering applications, notably in signal processing tasks such as wavelet-based image denoising \citep{Scott2006} and image registration \citep{ma2013robust, ma2015robust, yang2017remote}.

Some of the example methods we use to demonstrate our framework in Section \ref{sec:examples} have robust implementations.  These include the well-known robust multiple linear regression \citep{andrews1974robust, davies1993aspects, meng2013low, audibert2011robust, holland1977robust}, robust convex regression \citep{blanchet2019multivariate}, robust isotonic regression \citep{lim2018efficient, alvarez2012m}, and robust sparse regression \citep{SheOwen2011, Nguyen2013, alfons2013sparse, yang2018general, chang2018robust, ma2015robust}. The purpose of our experiments is not to compare the \LTE to each of these robust methods.  Rather, it is to demonstrate the flexibility and wide applicability of this computational framework and to show how it can obtain robust versions of existing non-robust implementations in lieu of developing on a case-by-case basis a robust version of a procedure from scratch.

Our framework's ability to simultaneously optimize over both the precision parameter and regression coefficients is a unique contribution to the literature.  To highlight this, we briefly discuss two lines of prior work listed above that are closely related to our proposed framework.

\subsubsection{Minimum distance estimators for sparse regression and image registration}

In the context of sparse regression, \citet{WanJiaHua2013} and \citet{Lozano2016} propose minimum distance estimators that coincide with our formulation when an $\ell_1$-norm sparsity promoting regularizer is used; \cite{Lozano2016} employ a slight modification of applying a log transform on the empirical minimum distance criterion. The key difference between these prior approaches and the framework we propose here is in how the precision parameter is estimated or determined. \citet{WanJiaHua2013} propose a hybrid block alternating scheme where the regression coefficients are estimated by minimizing the \LTE criterion with the precision parameter fixed and then the precision parameter is chosen to maximize efficiency subject to satisfying an asymptotic breakdown point of 1/2. Their procedure alternates between these two steps. We refer to this approach as ``hybrid" since the algorithm iterates are not minimizing a single objective function. Based on their simulation experiments, they state that their algorithm appears to converge within 1 to 3 steps but they do not provide a convergence proof. \citet{Lozano2016} treat the precision parameter as a hyper-parameter that can be selected via cross-validation. Nonetheless, for a fixed precision parameter, the algorithm that \citet{Lozano2016} propose does come with algorithmic guarantees.

Both \citet{WanJiaHua2013} and \citet{Lozano2016} require pre-specifying a grid of values for the precision parameter. A fine grid enables finding a better precision parameter at the cost of more computational effort. In our work, we estimate the regression coefficients {\em and} precision parameter by solving an optimization problem. Like \citet{WanJiaHua2013}, we also employ a block alternating algorithm, but unlike their approach, our approach is not hybrid and is kept completely within an optimization framework, enabling us to provide algorithmic convergence guarantees (See \Prop{convergence}). We will also see that our strategy can lead to better statistical performance in our simulation studies. Intuitively, we anticipate this since our strategy enables exploring the joint space of regression coefficients and precision parameter more comprehensively. The improved empirical performance also comes without a huge additional computational cost since the precision update step requires solving a univariate optimization problem -- a modest computational trade-off compared to solving multiple regression problems over a pre-specified grid of precision parameters.

In the context of image registration, \cite{ma2013robust, ma2015robust, yang2017remote} also employ minimum distance estimation to robustly fit a linear model. The main difference between their work and ours is in how the precision parameter is determined. They propose a deterministic annealing approach to choosing the precision parameter. They also solve an optimization problem to minimize the \LTE criterion with respect to the regression coefficients for a fixed precision parameter and then decrease the precision parameter a user-defined amount and then re-estimate the regression coefficients, alternating between updating the regression coefficient estimates and the precision parameter. Again a key question about this approach is whether or not the algorithm iterate sequence is guaranteed to converge. %It is unclear when to stop the alternating procedure.

\subsubsection{Trimmed estimators for high dimensional regression}

An alternative approach to obtaining robustness is to maximize a trimmed likelihood. \cite{alfons2013sparse} employ this approach for sparse robust multiple linear regression, namely they propose estimating a sparse regression coefficient vector $\V{\beta}$ by solving the following optimization problem
\begin{eqnarray}
\label{eq:trim}
\underset{\V{\beta}}{\min}\; \frac{1}{2} \sum_{i=1}^h \VE{r}{[i]}(\V{\beta})^2 + \lambda \lVert \V{\beta} \rVert_1,
\end{eqnarray}
where $\V{r}(\V{\beta}) = \V{y} - \M{X}\V{\beta}$ is a vector of residuals and $\VE{r}{[i]}(\V{\beta})$ is the $i$th order statistic of $\V{r}(\V{\beta})$. The nonnegative parameter $\lambda$ trades off model fit with sparsity in $\V{\beta}$. The trimming hyper-parameter $h$ imparts robustness to the standard residual sum of squares term by ``trimming away" observations with large residuals. \cite{yang2018general} extend the sparse trimmed least squares estimator introduced in \cite{alfons2013sparse} to a general framework for robust penalized estimation similar to ours in the sense that they introduce a single framework for computing structured robust regression problems.

The robustness of the estimator hinges on a good choice of $h$. \cite{alfons2013sparse} recommend using prior knowledge to set $h$ at a nominal value while \cite{yang2018general} employ cross-validation to select it in practice.

The hyper-parameter $h$ plays the same role as the precision parameter in the \LTE formulation.  Thus, a first key difference between the approach proposed by \cite{yang2018general} and ours is that we jointly estimate both the structured model and amount of trimming.  This approach has three benefits.  First, we reduce the potential for cross-validation to only any regularization parameters associated with the structure incentivizing penalties, e.g.\@ $\lambda$ in \Eqn{trim}.  Second, our framework enables a continuous (and therefore, larger) search space for choosing the precision parameter, as opposed to pre-specifying a finite but potentially very large grid of trimming parameters when there are many observations.  Third, our framework estimates both the regression coefficients and the precision parameter within an optimization framework, enabling convergence guarantees over the iterates.  %The convergence guarantees provide additional confidence in the estimated parameters than one can obtain from corresponding heuristic procedures.

A second key difference between the approach proposed by \cite{yang2018general} and ours is that the precision parameter in our framework performs a ``soft-trimming" action by adaptively choosing new down-weights for observations that are less consistent with the proposed model in each iteration.  Thus, rather than a single trim applied to all the observations, this enables additional flexibility for individual observations to contribute to the model fit varying amounts  and automatically updates the trimmed amount adaptively.  Section \ref{sec:robust_lasso} demonstrates the advantages of our approach.

\section{Robust regression with the \LT criterion}
\label{sec:L2E}

Let $f$ be the true but unknown density generating the observed data $y_{1}, \ldots, y_n \in \Real$, and let $\hat{f}_{\V{\theta}}$ be a probability density function indexed by a parameter $\V{\theta} \in \Theta \subset \Real^{q}$ that approximates $f$.  We assume throughout that all vectors are column vectors.  If we were to estimate $f$ using the $\hat{f}_{\V{\theta}}$ that is closest to it, we could minimize the L$_2$ distance between $f$ and $\hat{f}_{\V{\theta}}$ in lieu of the negative log-likelihood with
\begin{eqnarray}
\label{eq:ISE}
\min_{\Vhat{\theta} \in \Theta} \int \left [\hat{f}_{\V{\theta}}(y) - f(y) \right ]^2 dy.
\end{eqnarray}

In practice, however, we do not know $f$ and so identifying $\hat{\V{\theta}}$ in this way is impossible.  While we typically cannot minimize the L$_2$ distance between $f$ and its estimate $\hat{f}_{\V{\theta}}$ directly, we \emph{can} minimize an unbiased estimate of this distance.  To observe this, we first expand the quadratic integrand in \Eqn{ISE}, rewriting it as 
\begin{eqnarray*}
\label{eq:discreteL2E}
\int \hat{f}_{\V{\theta}}(y)^2 \,dy - 2\int \hat{f}_{\V{\theta}}(y)\,f(y)\,dy +  \int f(y)^2 \, dy.
\end{eqnarray*}
Notice that the second integral is the expectation $E_{Y}[\hat{f}_{\V{\theta}}(Y)]$, where $Y$ is a random variable drawn from $f$. Therefore, the sample mean provides an unbiased estimate of this quantity. Meanwhile, the third integral does not depend on $\V{\theta}$ so we can exclude it in the minimization.  In this way, we arrive at the the following fully data-based loss function $h(\V{\theta})$ that provides an unbiased estimate for \Eqn{ISE} up to an irrelevant additive constant
\begin{eqnarray}
\label{eq:l2e}
h(\V{\theta}) & = & \int \hat{f}_{\V{\theta}}(y)^2 dy - \frac{2}{n}\sum_{i=1}^n \hat{f}_{\V{\theta}}(y_i),
\end{eqnarray}
assuming $\hat{f}$ is square integrable over an appropriate region.  
Minimizing over this fully observed loss function presents us with our estimator $\Vhat{\theta}$, also called an \LTE \citep{Sco2001}.  We discuss how our computational framework provides intuition for how the \LTE imparts robustness in Section~\ref{sec:intuition}.

\subsection{Regression model formulation}\label{sec:model}

Let $\V{y} \in \Real$ denote a vector of $n$ observed responses and let $\M{X} \in \Real^{n \times p}$ denote the corresponding observed design matrix of $p$-dimensional covariates. The standard linear model assumes the response and covariates are related via the model
\begin{eqnarray*}
\V{y} & = & \M{X}\V{\beta}_{0} + \tau_{0}\Inv\V{\varepsilon},
\end{eqnarray*}
where $\V{\beta}_{0} \in \Real^p$ is an unobserved vector of regression coefficients, $\tau_{0} \in \Real_{+}$ is an unobserved precision parameter, and the unobserved noise $\VE{\varepsilon}{i} \in \Real$ for $1 \le i \le n$ are independently and identically distributed (iid) standard Gaussian random variables.  We phrase the regression model in terms of the precision rather than the variance to obtain a more straightforward optimization problem later.

Let $\V{\theta} = (\V{\beta}\Tra, \tau)\Tra$ denote the vector of unknown parameters.  Additionally, let $\V{r}$ denote the residual vector obtained from the current prediction estimate for $\V{\beta}$ so that its $i^{th}$ component is $\VE{r}{i} = \VE{y}{i} - \Mr{x}{i}\Tra \V{\beta}$, where $\Mr{X}{i} \in \Real^{p}$ is the vector containing the $i^{th}$ row of $\M{X}$.  Given any suitable pair of $\V{\beta}$ and $\tau$, the conditional density of $\VE{y}{i}$ for $1 \le i \le n$ is
\begin{eqnarray*}
\hat{f}^{(i)}_{\V{\theta}}(\VE{y}{i}) & = & \frac{\tau}{\sqrt{2\pi}} \exp\left(- \frac{\tau^2}{2} \VE{r}{i}^2 \right).
\end{eqnarray*}
As recommended in \citet{Sco2001}, when utilizing the \LTE loss function for linear regression, we average the L$_2$ distance over the observed data and minimize
\begin{eqnarray}
	\label{eq:l2e_unpenalized}
	h(\V{\theta}) =& \frac{1}{n} \sum_{i=1}^n h^{(i)}(\V{\theta}) =& \frac{\tau}{2\sqrt{\pi}} - \frac{\tau}{n}\sqrt{\frac{2}{\pi}} \sum_{i=1}^n \exp \left (- \frac{\tau^2}{2}\VE{r}{i}^2 \right),
\end{eqnarray}
where
\begin{eqnarray*}
h^{(i)}(\V{\theta})  \;=\;  \int_{-\infty}^\infty  \left [\hat{f}^{(i)}_{\V{\theta}}(\VE{y}{i}) \right ]^2 d\VE{y}{i} - 2 \; \hat{f}^{(i)}_{\V{\theta}}(\VE{y}{i}) 
 \;=\;  \frac{\tau}{2\sqrt{\pi}} - \tau\sqrt{\frac{2}{\pi}} \exp \left (- \frac{\tau^2}{2}\VE{r}{i}^2 \right).
\end{eqnarray*}
%As recommended in \citet{Sco2001}, when utilizing the \LTE loss function for linear regression, we average the L$_2$ distance over the observed data and minimize
%\begin{eqnarray}
%	\label{eq:l2e_unpenalized}
%	h(\V{\theta}) =& \frac{1}{n} \sum_{i=1}^n h^{(i)}(\V{\theta}) =& \frac{\tau}{2\sqrt{\pi}} - \frac{\tau}{n}\sqrt{\frac{2}{\pi}} \sum_{i=1}^n \exp \left (- \frac{\tau^2}{2}\VE{r}{i}^2 \right).
%\end{eqnarray}
The solution $\Vhat{\theta} = (\Vhat{\beta}\Tra, \hat{\tau})\Tra$ of \Eqn{l2e_unpenalized} contains the \LTE regression estimates.

\section{Computational framework}
\label{sec:computationalframework}
We pose our estimation and model fitting task as a non smooth optimization problem. We refer readers, who may be relatively new to non smooth optimization, to \cite{LangeChiZhou2014} and \cite{PolsonScottWillard2015} for brief overviews on optimization techniques for fitting models like the ones considered in this paper. For a broader and more comprehensive reference on optimization for fitting statistical models, we refer readers to \cite{lange2013optimization, lange2010numerical}.

We present a computational framework for performing robust structured regression using the L$_{2}$ criterion described in Section \ref{sec:L2E}.  We do this by introducing a general algorithm for combining the \LTE method \citep{Sco2001, Sco2009} with a general structural constraint or penalty term $\phi(\V{\beta})$. For example, suppose we wish to enforce a nonnegativity constraint on the regression coefficients $\V{\beta}$. Then we can take $\phi(\V{\beta}) = \iota_C(\V{\beta})$, the indicator function of the nonnegative orthant  $C = \{\V{\beta} \in \Real^p : \VE{\beta}{j} \geq 0, j = 1, \ldots, p\}$.
Recall that the indicator function of a set $C$, denoted $\iota_C(\V{\beta})$, is a function that takes values on the extended reals and is zero when $\V{\beta} \in C$ and is $\infty$ otherwise. As another example, $\phi(\V{\beta})$ may be an indicator function requiring that the elements of $\V{\beta}$ satisfy a monotonicity constraint. Other examples include taking $\phi(\V{\beta})$ to be sparsity inducing penalities like the $\ell_1$-norm \citep{tibshirani1996regression} or elastic net \citep{Zou2005}. Section \ref{sec:examples} contains several examples of potential constraint terms $\phi(\V{\beta})$.  Concretely, we seek a minimizer of the objective function
\begin{eqnarray}
\label{eq:objective}
	\ell(\V{\beta}, \tau) & = & h(\V{\beta}, \tau) + \phi(\V{\beta})
\end{eqnarray}
subject to $\V{\beta} \in \Real^{p}$ and $\tau \in [\tau_{\min}, \tau_{\max}]$, where $\tau_{\min} \in \Real$ and $\tau_{\max} \in \Real$ are minimum and maximum values for $\tau$, respectively.

There are two computational challenges in minimizing \Eqn{objective}. The first is that $\ell$ is non-convex in $\V{\theta}$ since $h(\V{\theta})$ is non-convex.  The second is that commonly used constraint terms $\phi(\V{\beta})$ are often non-smooth or non-differentiable.  We focus on the case where the $\phi$ are nonnegative, continuous, convex functions. Continuity and convexity ensures that $\phi$'s proximal mappings, a key algorithmic primitive to our framework, are well defined -- namely, they always exist and are unique.

In minimizing \Eqn{objective}, we utilize the key property that the block derivatives of $h$ with respect to $\V{\beta}$ and $\tau$, that is $\nabla_{\V{\beta}}h(\V{\beta}, \tau)$ and $\frac{\partial}{\partial \tau}h(\V{\beta}, \tau)$, respectively, are Lipschitz differentiable.

\begin{proposition} \label{prop:lipschitz} 
%	Given the assumptions in Section \ref{sec:model}, let $\sigma(\M{X})$ denote the largest singular value of the design matrix $\M{X}$, let $g(r) = (1 - \tau^2r)\exp(-\tau^2r^2/2)$, and let $r^\star = -  \frac{1}{2\tau^2}[\sqrt{1 + 4\tau^2} - 1]$.  
	The \LTE loss function $h(\V{\beta}, \tau)$ is block Lipschitz differentiable with respect to $\V{\beta}$ and $\tau$ so that
	\begin{eqnarray*}
	\lVert \nabla_{\V{\beta}}h(\V{\beta}, \tau)
	- \nabla_{\V{\beta}}h(\Vtilde{\beta}, \tau)
	\rVert_2 & \leq & 
	L_\beta(\tau) \lVert \V{\beta} - \Vtilde{\beta} \rVert_2
%	L_\beta \,\frac{\tau^3}{n}\,\sqrt{\frac{2}{\pi}}\,g(r^\star) \,\sigma(\M{X})^{2}\, \lVert \V{\beta} - \Vtilde{\beta} \rVert_2
	\end{eqnarray*}
	for all $\V{\beta}$ and $\Vtilde{\beta}$, and
	\begin{eqnarray*}
	\left \lvert \frac{\partial}{\partial \tau}h(\V{\beta},\tau) - \frac{\partial}{\partial \tau}h(\V{\beta},\tilde{\tau}) \right\rvert &\leq& L_\tau(\V{\beta}) \, \lvert \tau - \tilde{\tau} \rvert
	\end{eqnarray*}
	for all $\tau$ and $\tilde{\tau}$. The Lipschitz constant $L_\beta(\tau)$ is given by
	\begin{eqnarray*}
	L_\beta(\tau) & = & \frac{\tau^3}{n}\sqrt{\frac{2}{\pi}}\sigma(\M{X})^2,
	\end{eqnarray*}
	where $\sigma(\M{X})$ is the largest singular value of the design matrix $\M{X}$, 
	The Lipschitz constant $L_\tau(\V{\beta})$ is given by
	\begin{eqnarray*}
	L_\tau(\V{\beta}) & = & \frac{3}{n}\sqrt{\frac{2}{\pi}} \frac{\lVert \V{r} \rVert_{2}^2}{\rho} \exp\left( - \frac{1}{2} \right),
	\end{eqnarray*}
	where $\rho = \underset{i : r_i \neq 0}{\min}\; \lvert r_i \rvert$.
\end{proposition}
%\begin{proof}
	The proof is given in the supplement. The block Lipschitz differentiability of the \LTE criterion function $h(\V{\beta}, \tau)$ and the regularity conditions on $\phi$ lead us to employ a block coordinate descent algorithm to minimize \Eqn{objective}. At a high level, we alternate between minimizing with respect to $\V{\beta}$ holding $\tau$ fixed, and then minimizing with respect to $\tau$ holding $\V{\beta}$ fixed.  Therefore, at the $k^{th}$ update, we aim to solve the following two subproblems: \\
\noindent {\bf Subproblem 1: Update $\V{\beta}$}
\begin{eqnarray}
\label{eq:beta_update}
\Vn{\beta}{k} & =  & \underset{\V{\beta} \in \Real^{p}}{\arg\min}\; h(\V{\beta}, \tau^{(k-1)}) + \phi(\V{\beta}), \text{ and}
\end{eqnarray}
\noindent {\bf Subproblem 2: Update $\tau$}
\begin{eqnarray}
\label{eq:tau_update}
\tau^{(k)} & = & \underset{\tau \in [\tau_{\min}, \tau_{\max}]}{\arg\min}\; h(\Vn{\beta}{k}, \tau).
\end{eqnarray}

In practice, we cannot exactly solve either subproblem and instead take a few proximal gradient descent steps to partially minimize or inexactly solve \eqref{eq:beta_update} and \eqref{eq:tau_update}.  Note that the algorithm always makes progress towards minimizing the loss function, namely each update is guaranteed to monotonically decrease the overall loss function $\ell(\V{\theta})$, a feature that all block coordinate descent algorithms possess as a special case of majorization-minimization algorithms \citep{Lange2016}. Recall that proximal gradient descent is a first order iterative method for solving optimization problems of the form
\begin{eqnarray}
\label{eq:pg_objective}
\underset{\V{\theta}}{\text{minimize}}\; h(\V{\theta}) + \phi(\V{\theta}),
\end{eqnarray}
where $h$ is a Lipschitz differentiable function and $\phi$ is a convex and lower semicontinuous function \citep{Combettes2005, ParikhBoyd2014}. Further recall that the proximal map of $\phi$ is given by
\begin{eqnarray*}
\prox_{\phi}(\V{\theta}) & = & \underset{\Vtilde{\theta}}{\arg\min}\; \frac{1}{2} \lVert \Vtilde{\theta} - \V{\theta } \rVert_2^2 + \phi(\Vtilde{\theta}).
\end{eqnarray*}
The proximal map exists and is unique whenever $\phi(\V{\theta})$ is convex and lower semicontinuous. Many regularizers $\phi(\V{\beta})$ that are useful for recovering models with structure satisfy these conditions and also admit proximal maps that can be evaluated using either an explicit formula or an efficient algorithm. For example, the proximal map of the scaled $\ell_{1}$-norm $\lambda \lVert \cdot \rVert_{1}$ is the element-wise soft-thresholding operator, namely 
\begin{eqnarray}
\label{eq:lasso}
\left [\prox_{\lambda\lVert \cdot \rVert_{1}}(\V{\theta}) \right]_i & = & \text{sign}(\VE{\theta}{i})\max(\lvert \VE{\theta}{i}\rvert - \lambda,0).
\end{eqnarray}
The proximal map can be viewed as the generalization of the Euclidean projection, which we will refer to simply as the projection. Specifically, the projection of a point $\V{\theta}$ onto a set $C$ is the point $\mathcal{P}_C(\V{\theta}) \in C$ that is closest in Euclidean distance to $\V{\theta}$, namely
\begin{eqnarray*}
\mathcal{P}_C(\V{\theta}) & = & \underset{\Vtilde{\theta} \in C}{\arg\min}\; \lVert \Vtilde{\theta} - \V{\theta} \rVert_2.
\end{eqnarray*}
It is not hard to see that the proximal map of the indicator function $\iota_C$ of a set $C$ is the projection onto the set $C$ and moreover that the projection exists and is unique when $C$ is a closed convex set. For example, when $C = [\tau_{\min}, \tau_{\max}]$
\begin{eqnarray*}
\prox_{\iota_{[\tau_{\min}, \tau_{\max}]}}(\tau) & = & \mathcal{P}_{[\tau_{\min}, \tau_{\max}]}(\tau).
\end{eqnarray*}
As its name suggests, the proximal gradient descent method for solving problems of the form described in \Eqn{pg_objective} combines a gradient descent step with a proximal step. Given a current iterate $\V{\theta}$, the next iterate $\V{\theta}^+$ is computed as
\begin{eqnarray}
\label{eq:proximal_gradient}
\V{\theta}^+ & = & \prox_{t \phi}(\V{\theta} - t\nabla h(\V{\theta})),
\end{eqnarray}
where $t$ is a positive step-size parameter and $t\phi$ is the function $\phi$ scaled by $t$.

We emphasize that our framework does not require exactly computing the global minimizers in \eqref{eq:beta_update} and \eqref{eq:tau_update} at each iteration.  Nonetheless, in spite of inexactly solving \eqref{eq:beta_update} and \eqref{eq:tau_update}, we will see that the algorithm still comes with some convergence guarantees. 

\noindent{\bf Remark.}
Note that we do make the modestly stronger assumption that $\phi$ is continuous in order to establish the convergence guarantees. Assuming continuity is not restrictive as commonly employed, convex nonsmooth $\phi$ include norms, compositions of norms with linear mappings, and indicator functions of closed convex sets are continuous.

\subsection{A general algorithm for \LTE robust structured regression}
\label{sec:genalgorithm}

\Alg{L2E} presents pseudocode for minimizing \Eqn{objective} using inexact block coordinate descent. For the update step on $\tau$, the operator $\mathcal{P}_{[\tau_{\min}, \tau_{\max}]}$ denotes the projection onto $[\tau_{\min}, \tau_{\max}]$. When updating $\V{\beta}$ in \eqref{eq:beta_update} and $\tau$ in \eqref{eq:tau_update}, we take a fixed number of proximal gradient steps, $N_\beta$ and $N_\tau$ respectively, in \Eqn{proximal_gradient}. The gradients for updating $\V{\beta}$ and $\tau$ are given by
\begin{eqnarray}\label{eq:gradh}
\nabla_{\V{\beta}} h(\V{\beta}, \tau) & = & -\frac{\tau^3}{n}\sqrt{\frac{2}{\pi}} \M{X}\Tra\M{W}\V{r},
\end{eqnarray}
where $\M{W} \in \Real^{n \times n}$ is a diagonal matrix that depends on $\V{\beta}$ with $i^{th}$ diagonal entry 
\begin{eqnarray}\label{eq:weights}
	\ME{w}{ii} & = & \exp\left[-\frac{\tau^2}{2} \VE{r}{i}^2 \right],
\end{eqnarray}
and
\begin{eqnarray*}
\frac{\partial}{\partial \tau} h(\V{\beta}, \tau) & = & \frac{1}{2\sqrt{\pi}} - \frac{1}{n}\sqrt{\frac{2}{\pi}} \left [ \sum_{i=1}^n \ME{w}{ii}  \left (1 -  \tau^2\VE{r}{i}^2 \right) \right],
\end{eqnarray*}
respectively.

\Alg{L2E} has the following convergence guarantee. Recall that a point $\V{\theta} = (\V{\beta}\Tra, \tau)\Tra$ is a first order stationary point of a function $f(\V{\theta})$ if for all directions $\V{v}$, the directional derivative $f'(\V{\theta}; \V{v})$ of $f$ is nonnegative.
%\Prop{convergence} states the convergence property of \Alg{L2E} and also imparts guidance on the choice of the step sizes $t_\beta$ and $t_\tau$.  Under appropriate step-sizes $t_{\beta}$ and $t_{\tau}$ in \Eqn{beta_update} and \Eqn{tau_update}, \Alg{L2E} produces a monotonically decreasing sequence of objective function values and its iterates converge to a first order stationary point of the objective function in \Eqn{objective}. 

\begin{proposition}
\label{prop:convergence} 
For any choice of $N_{\beta}$ and $N_{\tau}$, under modest regularity conditions on \Eqn{objective} and step sizes $t_\beta = L_\beta(\tau)\Inv$ and $t_\tau =  L_\tau\Inv$, where $L_\beta(\tau)$ and $L_\tau(\V{\beta})$ are given by \Prop{lipschitz}, the sequence $(\Vn{\beta}{k}, \tau^{(k)})$ generated by \Alg{L2E} has at least one limit point and all limit points are first order stationary points of \Eqn{objective}. If there are finitely many first order stationary points of \Eqn{objective}, then the sequence $(\Vn{\beta}{k}, \tau^{(k)})$ generated by \Alg{L2E} will converge to one of them.
\end{proposition}
%\begin{proof}
Details on the regularity conditions and proof is given in the supplement. Before moving on from this convergence result about our algorithmic framework, we briefly comment on the assumption about the number of stationary points. Assuming that the \LTE objective $\ell(\V{\beta}, \tau)$ in \Eqn{objective} has finitely many first order stationary points may seem rather strong but a closer inspection of $h(\V{\beta}, \tau)$ suggests that this might not be unreasonable. A more in-depth exploration on this assumption is given in the supplement.

\begin{algorithm}[ht]
	\caption{Block coordinate descent for minimizing \Eqn{objective}}
	\label{alg:L2E}
	Initialize $\Vn{\beta}{0}, \tau^{(0)}$ and fix $N_{\beta}, N_{\tau}$
	\begin{algorithmic}[1]
		\STATE $k \gets 0$
		\REPEAT
		\STATE $t_\beta \gets L_\beta(\tau^{(k)})\Inv$ \COMMENT{Update $L_\beta(\tau^{(k)})$ via \Prop{lipschitz}}
		\STATE $\V{\beta} \gets \Vn{\beta}{k}$ \COMMENT{Update $\V{\beta}$ \Eqn{beta_update}}
		\FOR{$i = 1, \ldots, N_\beta$}
		\STATE $\V{\beta} \gets \prox_{t_\beta \phi}\bigg [\V{\beta} - t_\beta \nabla_{\V{\beta}} h(\V{\beta}, \tau^{(k)})\bigg]$
		\ENDFOR
		\STATE $\Vn{\beta}{k+1} \gets \V{\beta}$ 
		\STATE $t_\tau \gets L_\tau(\Vn{\beta}{k+1})\Inv$ \COMMENT{Update $L_\tau(\Vn{\beta}{k+1})$ via \Prop{lipschitz}}
		\STATE $\tau \gets \tau^{(k)}$ \COMMENT{Update $\tau$ \Eqn{tau_update}}
		\FOR{$i = 1, \ldots, N_\tau$}
		\STATE $\tau \gets \mathcal{P}_{[\tau_{\min}, \tau_{\max}]}\left [\tau - t_\tau \frac{\partial}{\partial \tau} h(\Vn{\beta}{k+1}, \tau)\right]$
		\ENDFOR
		\STATE $\tau^{(k+1)} \gets \tau$ 
		\STATE $k \gets k + 1$
		\UNTIL{convergence}
	\end{algorithmic}
\end{algorithm}

\subsection{Algorithmic intuition}
\label{sec:intuition}
We present a simple scenario illustrating intuition for \Alg{L2E}. This scenario applies directly to two examples that we will discuss in the next section, namely isotonic and convex regression. Let the design matrix $\M{X}$ be the $n \times n$ identity matrix $\M{I}_{n}$, and let the structural constraint $\phi(\V{\beta})$ be the indicator function of a closed and convex set $C$.  Therefore, $\phi(\V{\beta}) = \iota_C(\V{\beta})$ and is zero when $\V{\beta} \in C$, and $\infty$ otherwise.  This scenario results in simplifications to \Eqn{beta_update}.  In particular, the update rule for $\V{\beta}$ becomes
\begin{eqnarray*}
	\V{\beta}^+   \; = \; \mathcal{P}_C\left (\V{z} \right),
\end{eqnarray*}
where $\mathcal{P}_C\left (\V{z} \right)$ is the Euclidean projection of $\V{z}= 
\M{w}\V{y} + (\M{I} - \M{W})\V{\beta}$ onto $C$, and $\M{W}$ is a diagonal matrix with diagonal elements defined in \Eqn{weights}. 

We observe how the \LTE imparts robustness through the action of $\M{W}$. Consider $\V{z} \in \Real^{n}$ as a vector of pseudo-observations, where each element $\VE{z}{i}$ is a convex combination of $\VE{y}{i}$ and the current prediction $\VE{\beta}{i}$.  If the current residual $\VE{r}{i}$ is large compared to the current precision $\tau$, $\ME{w}{i}$ is small and the corresponding pseudo-observation $\VE{z}{i}$ resembles the current predicted value $\VE{\beta}{i}$.  Meanwhile, if the current residual $\VE{r}{i}$ is small compared to the current precision $\tau$, the corresponding pseudo-observation resembles the observed response $\VE{y}{i}$.

Therefore, the pseudo-observations impart the following algorithmic intuition.  Given an estimate $\Vtilde{\beta}$ of the regression coefficients, the algorithm performs constrained least squares regression using a pseudo-response $\V{z}$, whose entries are a convex combination of the entries of the observed response $\V{y}$ and the prediction $\Vtilde{\beta}$. Observations with large current residuals relative to the current precision, are essentially replaced by their predicted value.

Thus, the algorithm can fit a fraction of the observations very well while also accounting for outlying observations by replacing them with pseudo-response values that are more consistent with a model that fits the data.  Notice that the algorithm is oblivious to whether large residuals come from outliers in the response or in the predictor variables.  Consequently, it can handle outliers arising from either source or both.  

\subsection{Robustifying existing non-robust implementations}
\label{sec:plugandplay}

We end this section with a discussion detailing how one can employ this computational framework to automatically ``robustify" existing non-robust structured regression implementations that solve problems of the form
\begin{eqnarray*}
	\min_{\vbeta} \; \frac{1}{2}\| \vy - \mx\vbeta\|^{2}_{2} + \phi(\vbeta).
\end{eqnarray*}  
Specifically, we can utilize existing non-robust solvers to perform line 6 in Algorithm \ref{alg:L2E}.  Recall that line 6 performs the $\vbeta$ update with
\begin{eqnarray*}
	\V{\beta}^+   \; = \; \mathcal{P}_C\left (\V{z} \right) \quad\quad \text{ or } \quad\quad \vbeta^+ \;=\; \prox_{t_\beta \phi}(\vz)
\end{eqnarray*}
depending on whether $\phi$ is a projection operator or a more general proximal mapping.  In both cases, we perform this step by calling the existing non-robust solver and inputing $\vz$ in place of the original response $\vy$.  Our computation for $\vz$, however, depends on whether or not the design matrix is the identity.  

If the design matrix is the identity, such as in the cases of isotonic and convex regression, then $\vz$ in Algorithm \ref{alg:L2E} line 6 simplifies to $\V{z} \;=\; \M{w}\V{y} + (\M{I} - \M{W})\V{\beta}$, where $\mw$ is as described in \Eqn{weights}.  Therefore, we perform line 6 in Algorithm \ref{alg:L2E} by inputting $\vz \;=\; \M{w}\V{y} + (\M{I} - \M{W})\V{\beta}$ in place of $\vy$ into the existing non-robust solver.

If the design matrix is not the identity, such as in the case of Lasso regression, then $\vz$ in Algorithm \ref{alg:L2E} line 6 is the more complex $\V{z} = \V{\beta} - t_\beta \nabla_{\V{\beta}} h(\V{\beta}, \tau)$, where $\nabla_{\V{\beta}} h(\V{\beta}, \tau)$ is as described in \Eqn{gradh}.  Recall from Section \ref{sec:intuition} that the $\vbeta$ update involves solving a penalized or regularized least squares problem with an identity design matrix of the form
\begin{eqnarray*}
		\vbeta^+ \;\;=\;\; \prox_{t_\beta \phi}(\vz) & = & \underset{\Vtilde{\beta}}{\text{minimize}}\; \frac{1}{2}\lVert \V{z} - \mi\Vtilde{\beta} \rVert_2^2 + t_\beta \phi(\Vtilde{\beta}),
\end{eqnarray*}
where $\mi$ is an identity design matrix.  Therefore, we perform line 6 in Algorithm \ref{alg:L2E} by inputting $\vz \;=\; \V{\beta} - t_\beta \nabla_{\V{\beta}} h(\V{\beta}, \tau)$ in place of the $\vy$ and the identity matrix in place of $\mx$ into the existing non-robust solver. 

\subsection{Practical considerations}
\label{sec:considerations}

We now give some guidance on how to set hyperparameters in \Alg{L2E} in practice. The constraint set $[\tau_{\min}, \tau_{\max}]$ on $\tau$ is introduced primarily for technical reasons.  Namely, we employ it to establish the existence of a limit point for the algorithm iterate sequence.  In practice, however, we have not seen the constraints strongly influence performance. Nonetheless, it is possible to run into a numerical issue if $\tau_{\min}$ is set to zero. Specifically, it is possible that the gradient step in the $\tau$-update outputs a negative value which would then be projected to 0, which would result in the Lipschitz constant $L_\beta(\tau)$ being set to zero, which would in turn lead to an undefined step size $t_\beta$. To guard against such a possibility, we recommend setting $\tau_{\min}$ as follows. A conservative estimate of the standard deviation follows from assuming that there is no association between the response and covariates, namely take $\hat{\sigma} = \sqrt{\frac{1}{n-1}\sum_{i=1}^n (y_i - \overline{y})^2}$ and attribute all of the variation in the response $\V{y}$ to noise. We then take $\tau_{\min} = \hat{\sigma}\Inv$. As for the upper bound, taking $\tau_{\max}$ to be infinity does not seem to create any issues in practice.

A natural question is how to set $N_\beta$ and $N_\tau$, the number of inner iterations for updating $\V{\beta}$ and $\tau$ respectively, in \Alg{L2E}. Choosing these maximum iteration values too small or too large can lead to slow convergence. In our experience, setting $N_\beta = N_\tau = 1$ does not make sufficient progress in minimizing the objective functions in  \Eqn{beta_update} and \Eqn{tau_update}. Meanwhile, setting $N_\beta$ and $N_\tau$ to be a larger value such as $1{,}000$ leads to diminishing returns in minimizing the objective functions in \Eqn{beta_update} and \Eqn{tau_update}. In our experiments, we set $N_{\beta}$ and $N_{\tau}$ to be $100$ as it strikes a balance between these two extremes.

Finally, given the nonconvexity of the \LTE objective function in \Eqn{objective}, some thought to initialization for our algorithm is needed. We recommend the following simple ``null model" initialization strategy. When we have a non-identity design matrix $\M{X}$, similar to choosing $\tau_{\min}$, we assume there is no association between the response and covariates. So, we set the initial regression coefficient vector $\Vn{\beta}{0} = \V{0}$. When the design matrix $\M{X}$ is the identity, we set $\Vn{\beta}{0} = \overline{y}\V{1}$, namely the vector of all ones $\V{1}$ multiplied by the mean $\overline{y}$ of the response $\V{y}$. Regardless of whether we have covariates or not, we set the initial precision to be $\tau^{(0)} = \text{MAD}(\V{y})\Inv$, the reciprocal of the median absolute deviations of the response $\V{y}$. We use this initialization strategy in all our examples in the next section. We include a simulation study that provides some evidence that the output of \Alg{L2E} is also stable to perturbations in this initialization heuristic in the supplement.

% =================================================================================
% EXAMPLES
% =================================================================================
%\input{C4-examples}

\section{Examples of \LTE robust structured regression}
\label{sec:examples}

We demonstrate how the computational framework presented in Section \ref{sec:computationalframework} can perform a wide variety of robust structured regression methods with the L$_{2}$ criterion.  Our examples highlight how to incorporate existing non-robust structural regression solvers to ``robustify" existing implementations.  We refer to the estimates obtained from optimizing the maximum likelihood and the L$_{2}$ criterion as the MLE and L$_{2}$E, respectively.  Software for implementing robust structured regression via the $L_{2}$ criterion is available in the \verb|L2E| package for R and will be available on the Comprehensive R Archive Network (CRAN).

% ------------------------------------------------------------------------------
% Ex 1: Multivariate Linear Regression
% ------------------------------------------------------------------------------
\subsection{\LTE robust multiple linear regression}

We first demonstrate the most basic usage of our framework for multivariate \LTE regression \citep{Sco2001, Sco2009}, where $\phi(\V{\beta}) = 0$.  Let $\M{X} \in \Real^{n \times p}$ with $\rank(\mx) = p$.  
To illustrate this example, we utilize data from an Italian bank \citep{riani2014monitoring}.  The response $\vy \in \Real$ is the annual investment earnings for each of $n=1{,}949$ banking customers.  The design matrix $\mx$ contains quantitative measurements on each of $p=13$ bank services. 

Since $\phi(\V{\beta}) = 0$, $\prox_{t_{\beta}\phi}$ is simply the identity operation.  Therefore, Subproblem 1 for updating $\V{\beta}$ in \Eqn{beta_update} reduces to iteratively performing the following: 1) computing the current residuals, 2) updating the weights $\ME{w}{ii}$ in \Eqn{weights}, and 3) updating $\vbeta$ with the current residuals and the gradient described in Section \ref{sec:genalgorithm}.

\begin{figure}[h]
	\centering
	\begin{subfigure}{.47\textwidth}
		\centering
		\includegraphics[width=\linewidth]{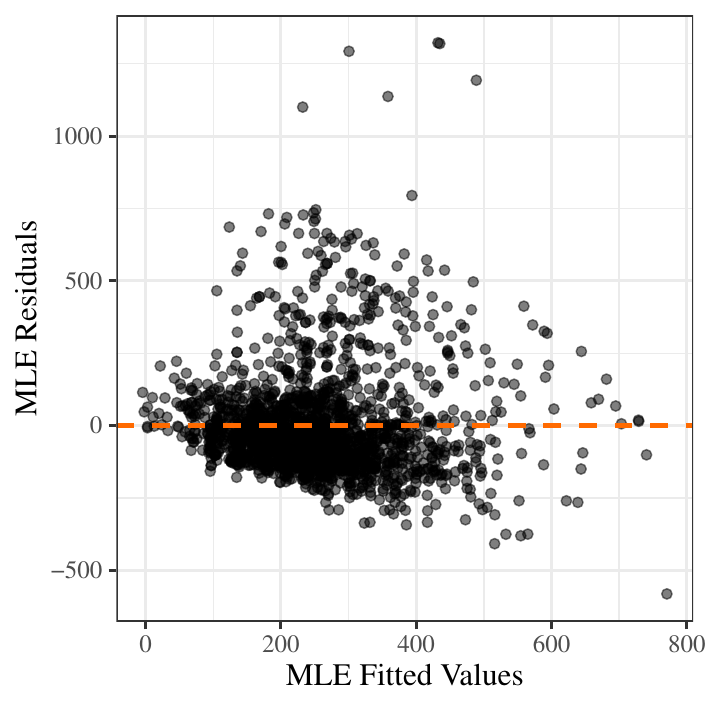}
		\caption{MLE fitted values vs. residuals.}
		\label{fig:ex1mleresids}
	\end{subfigure}\hspace{0.1in}
	\begin{subfigure}{.47\textwidth}
		\centering
		\includegraphics[width=\linewidth]{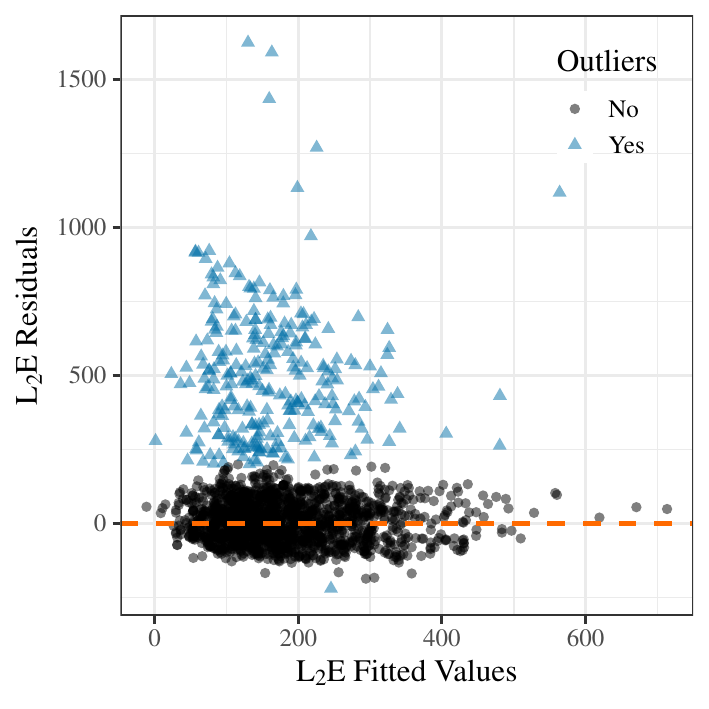}
		\caption{\LTE fitted values vs. residuals.}
		\label{fig:ex1l2eresids}
	\end{subfigure}
	\caption{\textit{(a) Asymmetric spread of points about the zero residual line suggests additional variation in the data not captured by the MLE linear fit. (b) Blue triangles denote outlying observations identified by the $L_{2}E$. Non-outlying observations are well-fit by the \LTE linear fit, as seen in normal distribution of points about the zero residual line.}}
	\label{fig:bankdataresids}
\end{figure}

Figures \ref{fig:ex1mleresids} and \ref{fig:ex1l2eresids} depict scatter plots of the fitted values against the residuals obtained with the MLE and L$_{2}$E, respectively.  A good fit is evidenced by normally distributed noise in the residuals -- namely, a symmetric scatter of points about the zero residual level (depicted by the horizontal orange dashed line).  \Fig{ex1mleresids}, however, shows a discernible pattern in the MLE residuals with asymmetric scatter of points about the zero residual level.  This indicates that additional trends in the data not captured by the Gaussian linear model remain in the residuals and are not captured by the MLE fit.

Meanwhile, \Fig{ex1l2eresids} shows that after excluding the outlying points identified by the automatic tuning of $\tau$ in our computational framework (depicted by the blue triangles), the residuals obtained using the \LTE fit are normally distributed about the zero residual level.  To identify outliers, one can compute the \LTE residuals and select those observations whose residuals exceed some factor of the precision parameter, e.g. 3 divided by $\tau$.  Thus, the \LTE adequately captures the linear relationship between investment earnings and bank services for the non-outlying customers.  Notice that \LTE regression can be recursively repeated on the outlying customers to identify an appropriate linear relationship between investment earnings and bank services for subgroups among the customers.

In addition to illustrating how the \LTE presents a more robust linear fit in the presence of outliers, this example also highlights another benefit of our computational framework.  Our framework enables the joint estimation of the regression coefficient vector $\V{\beta}$ and the precision $\tau$, enabling automatic identification of outlying observations in the data.  This is practically useful since the \LTE can simultaneously identify subpopulations within the data and appropriate fits for each of those groups when applied recursively to the subgroups.

% ------------------------------------------------------------------------------
% Ex 2: Isotonic regression
% ------------------------------------------------------------------------------
\subsection{\LTE robust isotonic regression}

We demonstrate how the computational framework proposed in Section \ref{sec:computationalframework} can perform robust isotonic regression using the L$_{2}$ criterion.  
Let an observed response $\V{y} \in \Real^{n}$ consist of $n$ samples drawn from a monotonic function $f$ sampled at discrete time points $t_{1} \leq t_2 \leq \dots \leq t_{n}$ with additive independent Gaussian noise.  We can express the $i^{th}$ entry of $\vy$ as
\begin{eqnarray*}
	\VE{y}{i} = f(t_{i}) + \epsilon_{i} \quad \text{ for } \quad 1 \le i \le n,
\end{eqnarray*}
where $f$ is monotonic, $\epsilon_{i} \overset{\text{iid}}{\sim} N(0,\frac{1}{\tau})$, and $\tau \in \Real_{+}$.  The goal of isotonic regression \citep{barlow1972statistical, barlow1972isotonic, isotone, lee1981quadratic, dykstra1982algorithm} is to estimate $f$ by solving
\begin{align*}
		& \min_{\beta(t_1),\, \dots\, ,\, \beta(t_n)}  \sum_{i=1}^{n} [\VE{y}{i} - \beta(t_{i})]^{2} \\
		& \text{subject to } \quad \beta(t_{1}) \le \beta(t_{2}) \le \cdots \le \beta(t_{n}).
\end{align*}
We then construct a piece-wise constant estimate for $f$ using the elements of the estimator $\hat{\V{\beta}} = \begin{pmatrix}\hat{\beta}(t_{1}) & \hat{\beta}(t_{2}) & \cdots & \hat{\beta}(t_{n}) \end{pmatrix}\Tra$.

For the corresponding \LTE problem, the design matrix $\M{X}$ is the $n \times n$ identity matrix $\M{I}_{n}$ and $\phi(\V{\beta}) = \iota_{\mathcal{M}}(\V{\beta})$ is the indicator function over the set of vectors $\mathcal{M}$ satisfying element-wise monotonicity so that $\VE{\beta}{1} \le \VE{\beta}{2} \le \cdots \le \VE{\beta}{n}$ for $\V{\beta} \in \Real^{n}$.  Subproblem 1 for updating $\V{\beta}$ in \Eqn{beta_update} reduces to iteratively performing the following: 1) computing the current residuals, 2) updating the weights $\ME{w}{ii}$ in \Eqn{weights}, and then 3) updating $\vbeta$ using the current residuals and the gradient described in Section \ref{sec:genalgorithm} and projecting onto the set $\mathcal{M}$.  The \verb|gpava| function for implementing the generalized pool-adjacent-violators algorithm (generalized PAVA) in the \verb|isotone| package \citep{isotone} for R performs this last step.  Therefore, we harness a readily available non-robust isotone regression solver in line 6 of Algorithm \ref{alg:L2E}.

\begin{figure}[h]
	\centering
	\begin{subfigure}{.45\textwidth}
		\centering
		\includegraphics[width=\linewidth]{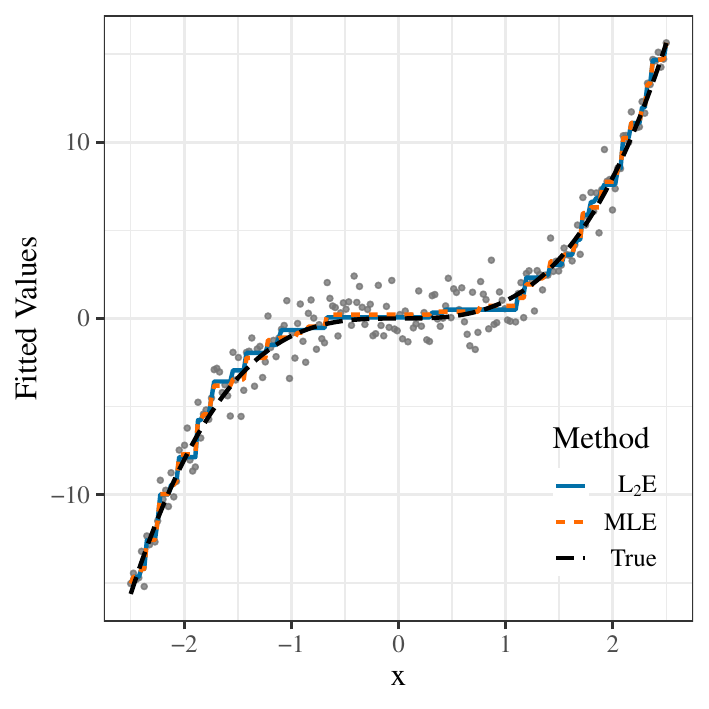}
		\caption{Isotonic regression without outliers.}
		\label{fig:isocleansol}
	\end{subfigure}\hspace{0.1in}
	\begin{subfigure}{.45\textwidth}
		\centering
		\includegraphics[width=\linewidth]{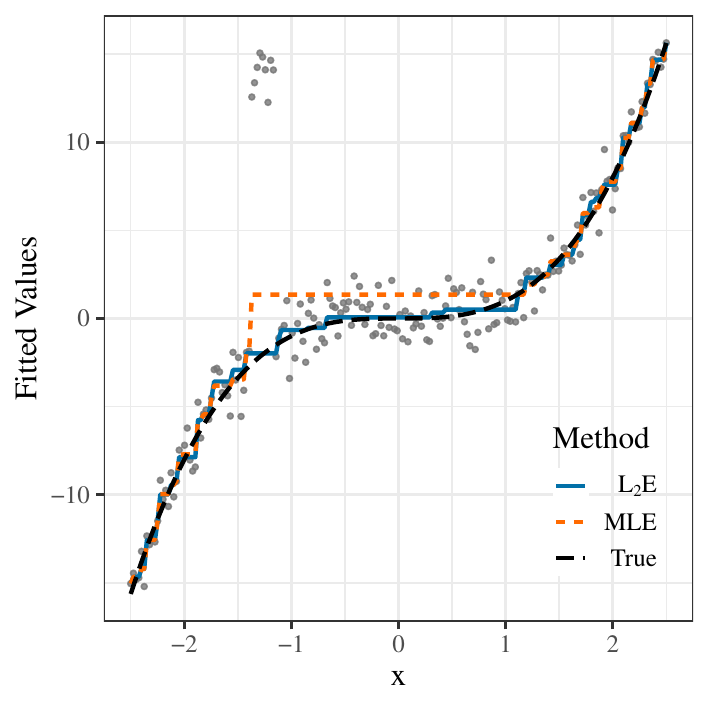}
		\caption{Isotonic regression with outliers.}
		\label{fig:isocontaminatedsol}
	\end{subfigure}
	\caption{\textit{The black, orange, and blue lines depict the true, MLE, and \LTE fits for isotonic regression, respectively.  The MLE and \LTE produce similar results in the absence of outliers.  The MLE is skewed towards the outliers while the \LTE provides a more robust estimate.}}
	\label{fig:isotonic}
\end{figure}

We illustrate with a univariate cubic function.  
\Fig{isocleansol} shows how the MLE and \LTE produce similar estimates in the absence of outliers.  The true underlying cubit fit $f$ is shown in black and the gray points depict the observations generated from $f$ with additive Gaussian noise.  The dashed orange line depicts the MLE obtained using generalized PAVA while the solid blue line depicts the \LTE.  Meanwhile, \Fig{isocontaminatedsol} shows how the MLE is skewed towards the outliers while the \LTE estimate remains less sensitive to them.

\begin{figure}[H]
	\centering
	\begin{subfigure}{.45\textwidth}
		\centering
		\includegraphics[width=\linewidth]{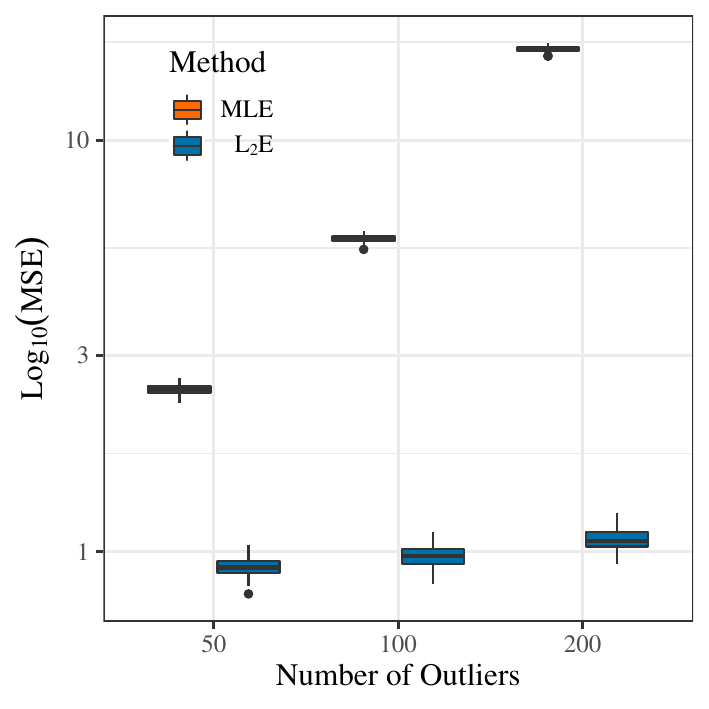}
		\caption{Isotonic regression mean squared error.}
		\label{fig:iso_mse}
	\end{subfigure}\hspace{0.1in}
	\begin{subfigure}{.45\textwidth}
		\centering
		\includegraphics[width=\linewidth]{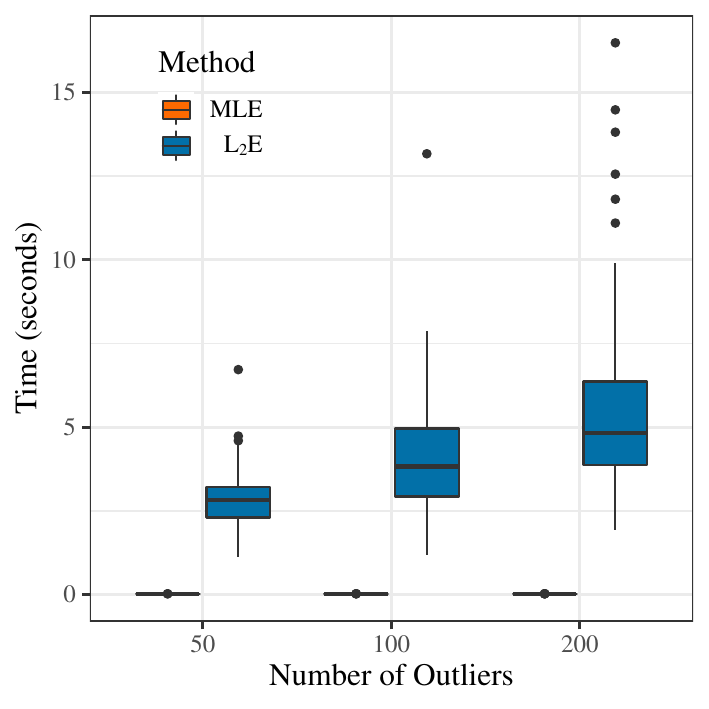}
		\caption{Isotonic regression time (in seconds).}
		\label{fig:iso_time}
	\end{subfigure}
	\caption{\textit{Monte Carlo experiments for isotonic regression with $n=1{,}000$ observations drawn from a univariate cubic function with additive Gaussian noise.  Boxplots depict the (a) mean squared error (MSE) and (b) time in seconds required for $50, 100,$ and $200$ outliers over $100$ replicates for the MLE in orange and the \LTE in blue.  Unsurprisingly, the \LTE requires more time since its solution requires multiple computations of the MLE.  Parts (a) and (b) highlight the trade-off between MSE and time between the MLE and \LTE solutions.
	}}
	\label{fig:isotonic_experiments}
\end{figure}

\Fig{isotonic_experiments} depicts results of Monte Carlo simulations comparing the MLE and the \LTE while varying the number of outliers.  We simulate three datasets with $n=1{,}000$ observations of a cubic function with additive Gaussian noise and $50, 100$, and $200$ outliers, respectively.  We introduce outliers by selecting points from approximately the $25^{th}$ quartile along the x-axis and assigning them a value equal to slightly less than the maximal polynomial value and additive standard Gaussian noise.  This corresponds to simulating samples from a bimodal distribution to create high leverage points in the covariate space.  We employ the \verb|gpava| function in the \verb|isotone| package \citep{isotone} for R to obtain the MLE.  We obtain $100$ replicates for each scenario on a 3.00 GHz Intel Core i7 computer with 32 GB of RAM and present boxplots of the mean squared error (MSE) and time in seconds.  We obtain the MSE between the model $\vy$ and the computed solution.

The MLE produces increasingly larger MSE as the number of outliers increases.  Meanwhile, the \LTE produces a much smaller increase in MSE for the same number of outliers but requires more time in seconds since the \LTE solution employs multiple computations of the MLE procedure.  Thus, the \LTE can produce an isotonic regression fit that is much less sensitive to outliers than the MLE.  This example highlights how our framework can utilize a readily available non-robust isotonic regression solver to automatically perform robust isotonic regression using the L$_{2}$ criterion.

% ------------------------------------------------------------------------------
% Ex 3: Convex regression
% ------------------------------------------------------------------------------
\subsection{\LTE robust convex regression}

We demonstrate how the computational framework proposed in Section \ref{sec:computationalframework} can perform robust convex regression using the L$_{2}$ criterion. 
For illustrative purposes, we consider the univariate case \citep{wang2012shape, ghosal2017univariate}.  However, applying our framework to multivariate convex regression \citep{seijo2011nonparametric, mazumder2019computational, guntuboyina2015global, birke2007estimating, lim2012consistency, hannah2013multivariate, meyer2003test, Bertsimas2021, ChenMazumder2021, AybatWang2016, LinSunToh2020} can be performed in a similar manner.  
Let an observed response $\V{y} \in \Real^{n}$ consist of $n$ samples drawn from a convex function $f$ sampled at discrete time points $t_{1} \leq t_2 \leq  \dots \leq t_{n}$ with additive independent Gaussian noise.  We can express the $i^{th}$ entry of $\vy$ as
\begin{eqnarray*}
	y_{i} = f(t_{i}) + \epsilon_{i} \quad \text{ for } \quad 1\le i \le n,
\end{eqnarray*}
where $f$ is convex, $\epsilon_{i} \overset{\text{iid}}{\sim} N(0,\frac{1}{\tau})$, and $\tau \in \Real_{+}$.  The goal of shape-restricted convex regression is to estimate $f$ by solving
\begin{align*}
	& \min_{\beta(t_1),\, \dots\,,\,  \beta(t_n)}  \sum_{i=1}^{n} [y_{i} - \beta(t_{i})]^{2} \\
	& \text{subject to } \quad \beta(t_{i}) \le
	\frac{t_{i+1}-t_{i}}{t_{i+1} - t_{i-1}} \beta(t_{i-1}) + \frac{t_{i}-t_{i-1}}{t_{i+1}-t_{i-1}} \beta(t_{i+1}) \quad \text{ for } \quad 2 \le i \le n-1.
\end{align*}
Let $\V{\beta} = \begin{pmatrix} \beta(t_{1}) & \beta(t_{2}) & \cdots & \beta(t_{n}) \end{pmatrix}\Tra$.  We can recast this constraint in terms of a scaled second-order differencing matrix $\M{D} \in \Real^{n \times n}$ with $\M{D}\V{\beta} \ge \V{0}$ so that all the elements of $\M{D}\V{\beta}$ are non-negative.  
We then construct a piece-wise constant estimate for $f$ using the elements of the estimator $\hat{\V{\beta}} = \begin{pmatrix}\hat{\beta}(t_{1}) & \hat{\beta}(t_{2}) & \cdots & \hat{\beta}(t_{n}) \end{pmatrix}\Tra$.

For the corresponding \LTE problem, the design matrix $\M{X}$ is the $n \times n$ identity matrix $\M{I}_{n}$ and $\phi(\V{\beta}) = \iota_{\mathcal{C}}(\V{\beta})$ is the indicator function over the set of vectors in $\mathcal{C} \equiv \{\V{\beta}: \M{D}\V{\beta} \ge {\bf 0} \}$.  Subproblem 1 for updating $\V{\beta}$ in \Eqn{beta_update} reduces to iteratively performing the following: 1) computing the current residuals, 2) updating the weights $\ME{w}{ii}$ in \Eqn{weights}, and then 3) updating $\vbeta$ using the current residuals and the gradient described in Section \ref{sec:genalgorithm} and projecting onto the convex cone $\mathcal{C}$.  The \verb|conreg| function in the \verb|cobs| package \citep{cobs} for R can be used to perform this last step.  Therefore, we can utilize a readily available non-robust convex regression solver in line 6 of Algorithm \ref{alg:L2E}.

\begin{figure}[h]
	\centering
	\begin{subfigure}{.45\textwidth}
		\centering
		\includegraphics[width=\linewidth]{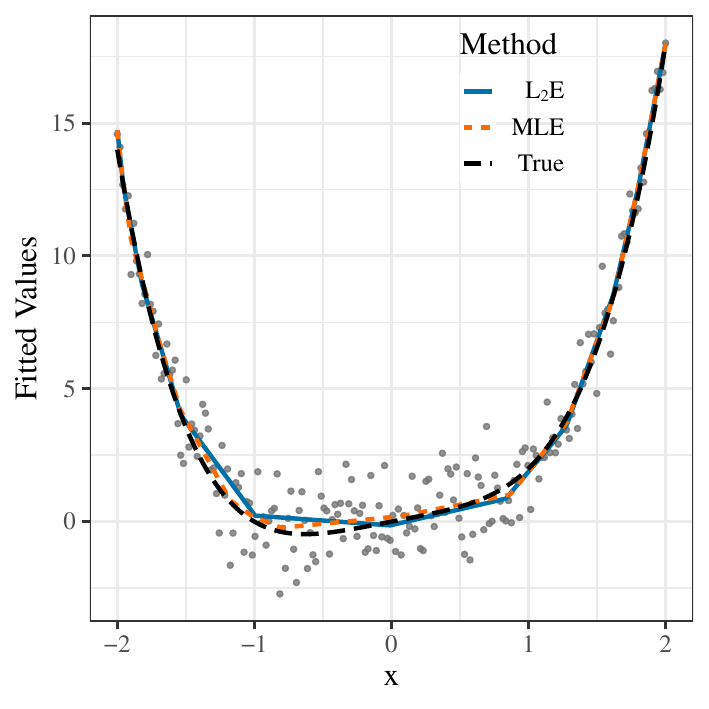}
		\caption{Convex regression without outliers.}
		\label{fig:convexcleansol}
	\end{subfigure}\hspace{0.1in}
	\begin{subfigure}{.45\textwidth}
		\centering
		\includegraphics[width=\linewidth]{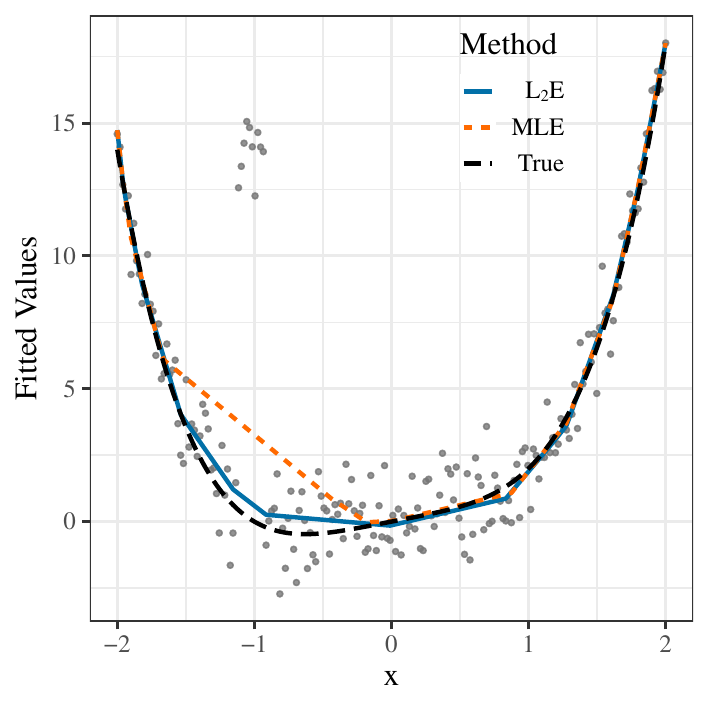}
		\caption{Convex regression with outliers.}
		\label{fig:convexcontaminatedsol}
	\end{subfigure}
	\caption{\textit{The black, orange, and blue lines depict the true, MLE, and \LTE fits for convex regression, respectively.  The MLE and \LTE produce similar results in the absence of outliers.  The \LTE is much less sensitive to outliers than the MLE.}}
	\label{fig:convex}
\end{figure}

\Fig{convexcleansol} shows how the MLE and \LTE produce similar fits in the absence of outliers.  The true underlying convex fit $f$ is shown in black and the gray points depict the observations generated from the true fit and some additive Gaussian noise.  The dashed orange line depicts the MLE obtained using the \verb|cobs| package in R while the solid blue line depicts the \LTE.  Meanwhile, \Fig{convexcontaminatedsol} shows how the MLE is substantially skewed towards the outliers while the \LTE is less distorted.  This example again highlights how the \LTE is less sensitive to outliers than the MLE.

\begin{figure}[H]
	\centering
	\begin{subfigure}{.45\textwidth}
		\centering
		\includegraphics[width=\linewidth]{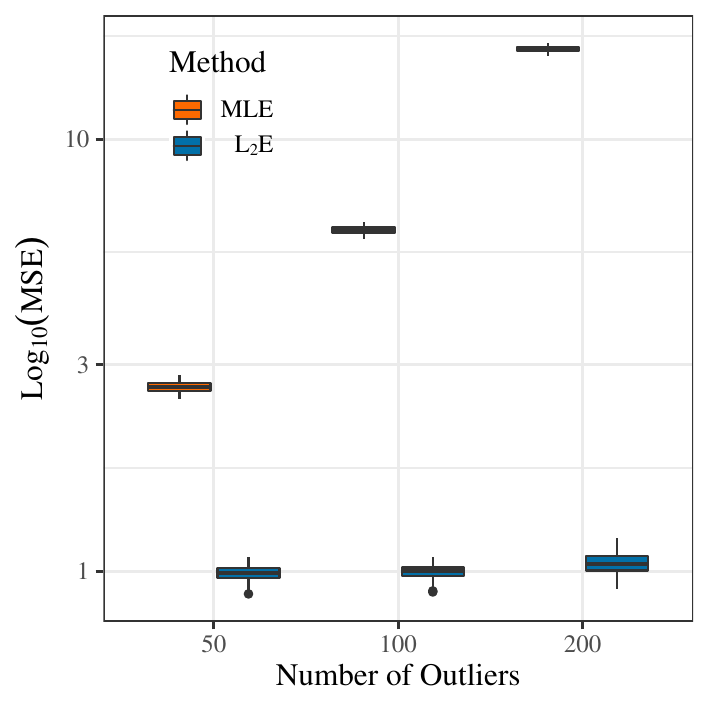}
		\caption{Convex regression mean squared error.}
		\label{fig:convex_mse}
	\end{subfigure}\hspace{0.1in}
	\begin{subfigure}{.45\textwidth}
		\centering
		\includegraphics[width=\linewidth]{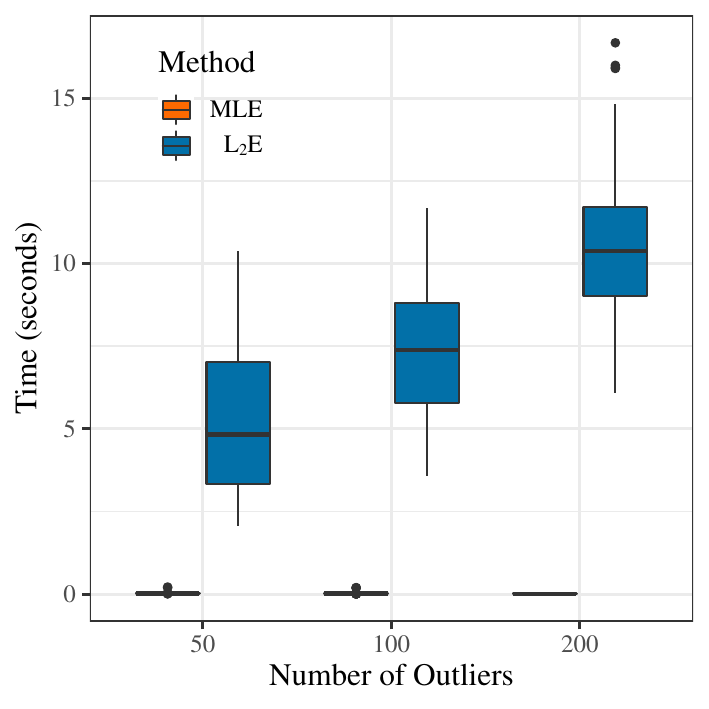}
		\caption{Convex regression time (in seconds).}
		\label{fig:convex_time}
	\end{subfigure}
	\caption{\textit{Monte Carlo experiments for convex regression with $n=1{,}000$ observations drawn from a convex function with additive Gaussian noise.  Boxplots of the (a) mean squared error (MSE) and (b) time in seconds required for $50, 100,$ and $200$ outliers over $100$ replicates are shown with the MLE in orange and the \LTE in blue.  Unsurprisingly, the \LTE requires more time as its solution employs multiple computations of the MLE.  Parts (a) and (b) highlight the trade-off between MSE and time between the MLE and \LTE solutions.}}
	\label{fig:convex_experiments}
\end{figure}

\Fig{convex_experiments} depicts the results of Monte Carlo simulations comparing the MLE and \LTE on shape-restricted convex regression while varying the number of outliers.  We simulate three datasets with $n=1{,}000$ observations using a fourth-order polynomial with additive Gaussian noise and $50, 100$, and $200$ outliers, respectively.  We introduce outliers by selecting points from approximately the $25^{th}$ quartile along the x-axis and assigning them a value that is equal to a little less than the maximal polynomial value and additive standard Gaussian noise.  This corresponds to simulating samples from a bimodal distribution to create high leverage points in the covariate space.  We employ the \verb|conreg| function in the \verb|cobs| package for R \citep{cobs} to obtain the MLE.  We obtain $100$ replicates on a 3.00 GHz Intel Core i7 computer with 32 GB of RAM and present boxplots of the MSE and time in seconds.

\Fig{convex_mse} highlights how the MLE produces increasingly larger MSE values as the number of outliers increases.  Meanwhile, the MSE obtained using the \LTE also grows slightly as the number of outliers increases but it is much less sensitive to outliers.  This example again underscores how the computational framework presented in Section \ref{sec:computationalframework} can perform a robust version of a structured regression problem utilizing a readily available non-robust implementation.

\subsection{\LTE robust $\ell_{1}$ penalized regression}
\label{sec:robust_lasso}

We demonstrate how the computational framework proposed in Section \ref{sec:computationalframework} can perform $\ell_{1}$ penalized regression via the L$_{2}$ criterion.  We utilize the Lasso \citep{tibshirani1996regression}
\begin{eqnarray*}
	\min_{\vbeta}\; \frac{1}{2} \lVert \vy - \mx\vbeta\rVert_{2}^{2} + \lambda \lVert \V{\beta} \rVert_1
\end{eqnarray*}
as our reference.  For the corresponding \LTE problem, let $\M{X} \in \Real^{n \times p}$ with $\rank(\mx) = p$ and let $\phi(\V{\beta}) = \lambda\lVert \V{\beta} \rVert_{1}$.  Subproblem 1 for updating $\V{\beta}$ in \Eqn{beta_update} reduces to iteratively performing the following: 1) computing the current residuals, 2) updating the weights $\ME{w}{ii}$ in \Eqn{weights}, and then 3) updating $\vbeta$ using the current residuals and the gradient described in Section \ref{sec:genalgorithm} and applying the element-wise soft-thresholding operator in \Eqn{lasso}.

We illustrate with real data on patients with prostate cancer from \citet{prostate}.  The response $\vy \in \Real^n$ is the percent of Gleason score (a measure of a prostate-specific antigen) for each of $n=97$ patients who were to receive a radical prostatectomy.  The design matrix $\mx$ contains quantitative measurements on each of $p=8$ clinical variables.  To introduce outliers in the covariates, we identify the top five percent of observations in $\mx$ with highest leverage and scale the entries in these five largest leverage points by $3.3$.  

\begin{figure}[H]
	\centering
	\scalebox{.9}{\includegraphics[width=\linewidth]{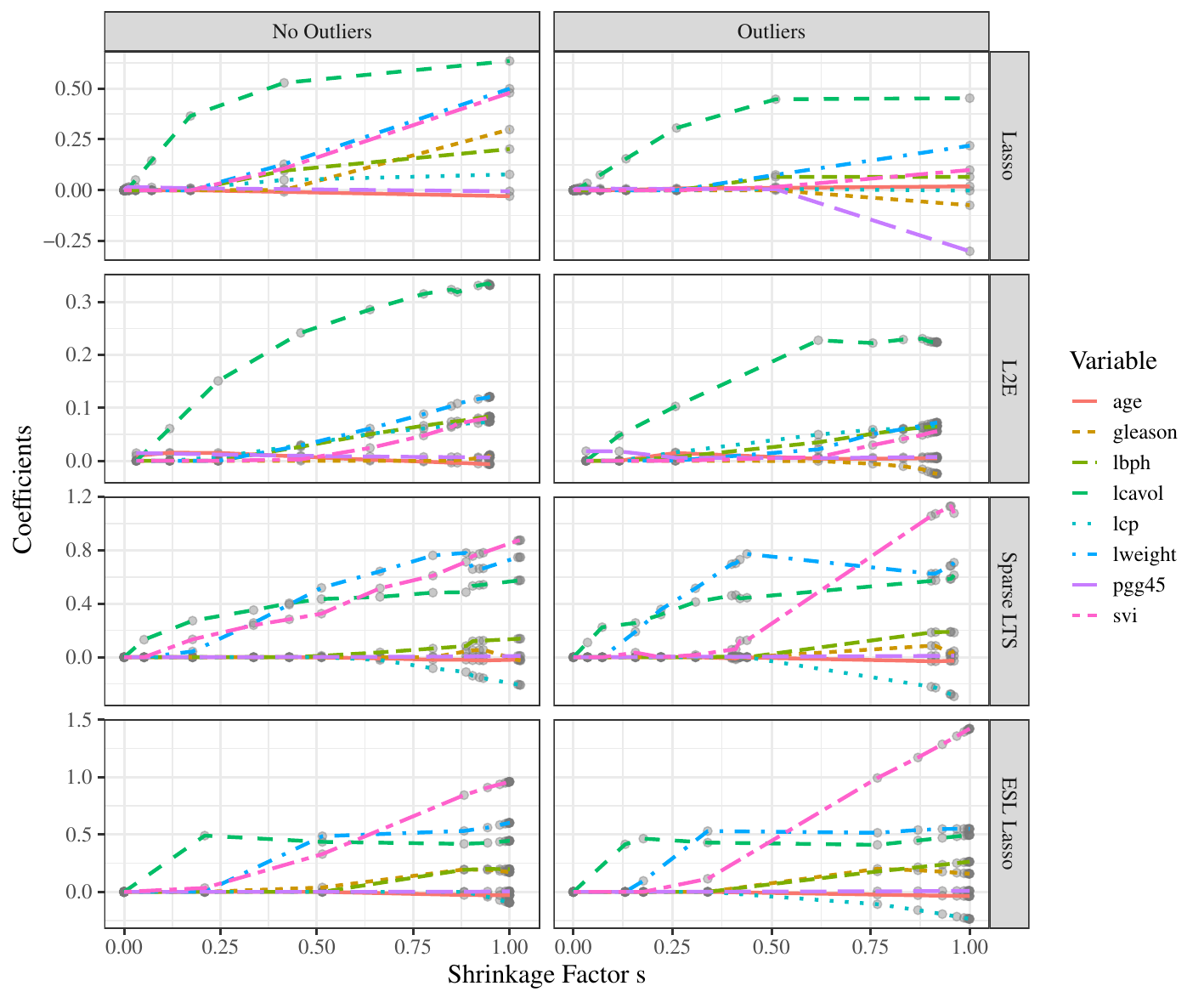}}
	\caption{\textit{Solution paths for Lasso, \LTE $\ell_{1}$ penalized regression, sparse least trimmed squares (Sparse LTS), and exponential squared loss Lasso (ESL Lasso) as a function of the shrinkage factor $s=\lVert \vbeta(\lambda)\rVert_{1}/ \lVert\vhbeta_{0}\rVert_{1}$, where $\vhbeta_{0}$ denotes the $\vbeta$ estimate obtained at $\lambda=0$ for each method.  Lasso and \LTE have similar interpretation in the absence of outliers and the \LTE solution suffers less distortion in the presence of outliers.}}
	\label{fig:sparse}
\end{figure}

Figure \ref{fig:sparse} depicts solution paths for Lasso, \LTE $\ell_{1}$ penalized regression, sparse least trimmed squares (Sparse LTS) \citep{yang2018general, alfons2013sparse}, and exponential squared loss Lasso (ESL Lasso) \citep{WanJiaHua2013} as a function of the shrinkage factor $s=\lVert \vbeta(\lambda)\rVert_{1}/ \lVert\vhbeta_{0}\rVert_{1}$.  We set $\vhbeta_{0}$ as the $\vbeta$ estimate obtained at $\lambda=0$ for each method and employ a $\lambda$ sequence with a log linear scale of $15$ values between $10^{-5}$ and a conservative data-dependent estimate of $\lambda$ at which $\vhbeta(\lambda) = {\bf 0}$.

Since all four methods employ the $\ell_{1}$ penalty, one can view the latter three methods as different approaches to obtaining robust Lasso.  Therefore, in the absence of outliers, we view the Lasso solution paths (top-left panel) as the control.  As a method of variable selection, the Lasso solution paths quantify the relative contributions of the covariates to the regression model.  Ideally, a robust implementation preserves these relative contributions in the presence of outliers.  Qualitatively, the \LTE solution paths most closely resemble the Lasso solution paths and suffers the least distortion in the presence of outliers.  By comparison, the Sparse LTS and ESL Lasso qualitatively look quite different from the Lasso solution paths, even in the absence of outliers.  For Sparse LTS, we employ the default trimming percentage (retains 75 percent of the data), so it should be robust to the five percent of outliers in the data. 

To present more quantitative comparisons, we additionally perform numerical experiments with synthetic data comprising a response $\vy \in \mathbb{R}^{100}$ and design $\mx \in \mathbb{R}^{100 \times 50}$ containing outliers in both the response and the covariates.  The entries of $\mx$ are iid standard Gaussian random variables and $\vy = \mx\V{\beta}_{0} + \V{\epsilon}$, where the first five entries of $\V{\beta}_{0}$ are equal to $1$ and the remaining entries are equal to $0$.  The entries of $\V{\epsilon}$ are also iid standard Gaussian random variables.  To introduce outliers in the response, we obtain $30$ percent additive contamination by shifting the first $30$ entries in $\vy$ by $5$.  Similarly, we introduce $30$ percent poor leverage points by shifting the entries in the first $30$ rows of $\mx$ by $5$.  

We employ 10-fold cross-validation for all methods to choose $\lambda$ employing a fine grid log linear scale of $100$ values between $10^{-8}$ and a conservative data-dependent estimate of $\lambda$ at which $\vhbeta(\lambda) = {\bf 0}$.  For \LTE and ESL Lasso, we cross-validate over $\lambda$ with respect to the \LTE objective in \Eqn{l2e_unpenalized}.  For Sparse LTS, we cross-validate over both $\lambda$ and the trimming parameter $h$ varying from $50$ to $90$ in increments of $5$, and evaluate performance via the median squared residuals in the hold-out fold.  We take the median instead of the average to account for the possibility of outliers in the hold-out fold.

Table \ref{table:sparse_experiments_v2} depicts the mean (and standard deviation) relative error compared with $\V{\beta}_{0}$, number of true positives, and number of false positives from L$_{2}$E, Sparse LTS, and ESL Lasso over 20 replicates.  \LTE obtains lower relative error on average and additionally selects fewer false positives.  Although Sparse LTS and ESL Lasso employ the $\ell_{1}$ penalty for variable selection, they both select nearly all the variables in the presence of outliers.

\begin{table}[ht]
	\centering
	\begin{tabular}{lrrrrrr}
		\hline
		Method & Rel. Error   && True Pos. &&   False Pos. &  \\ 
		\hline
		\LTE & 0.64 & (0.10) & 4.85 & (0.37) & 13.10 & (5.66) \\ 
		Sparse LTS & 1.13 & (0.23) & 5.00 & (0.00) & 44.95 & (0.22) \\ 
		ESL Lasso & 0.81 & (0.14) & 5.00 & (0.00) & 44.00 & (0.92) \\ 
		\hline
	\end{tabular}
	\caption{\textit{Mean (and standard deviation) of the relative error, number of true positives, and number of false positives obtained by L$_{2}$E, sparse least trimmed squares (Sparse LTS), and exponential squared loss Lasso (ESL Lasso) over $20$ replicates on $n=100$ and $p=50$ synthetic data with outliers in both the response and the covariates.  \LTE obtains lower relative error on average and additionally selects fewer false positives in variable selection.}}
	\label{table:sparse_experiments_v2}
\end{table}

%While there are several robust implementations of sparse regression and its variations \citep{alfons2013sparse, yang2018general, chang2018robust}, the main purpose of this example is not to compare \LTE $\ell_{1}$ penalized regression with those implementations.  Rather, it is to illustrate how the framework we present can wrap around existing maximum penalized estimators to automatically compute robust versions of those methods.

%\input{C5-discussion}

\section{Discussion}
\label{sec:discussion}

Least squares regression models can be extended to encode a wide array of prior structure through non smooth penalties and constraints. While regression via least squares -- and its constrained and penalized extensions -- does not require any parametric assumptions, making a normality assumption on the residuals opens the door to applying the \LTE method for robustly fitting a parametric regression model. In this work, we introduce a user-friendly computational framework, or recipe, for performing a wide variety of robust structured regression methods by minimizing the L$_{2}$ criterion.  We highlight that our framework can ``robustify" existing structured regression solvers by utilizing existing non-robust solvers in the $\V{\beta}$-update step in a plug-and-play manner.  Thus, our framework can readily incorporate newer and improved technologies for existing structured regression methods; as faster and better algorithms for these non-robust structured regression solvers appear, users may simply replace the previous solver with the new one in the $\V{\beta}$-update step.

We also highlight the significance of the convergence properties of our computational framework.  As long as the structural constraints or penalties satisfy convexity and continuity conditions, a solution obtained with our framework is guaranteed to converge to a first order stationary point.  Since many commonly-used structural constraints and penalties satisfy these conditions, our framework provides convergence guarantees for robust versions of many non-robust methods with readily available software.

We close by noting that our \LTE framework focuses on structured regression problems under a normality assumption, which may not be appropriate in all situations. Meanwhile, the \LTE framework has also been used to robustly estimate parametric models under different distributional assumptions, e.g.\@ Weibul \citep{Yang2013}, Poisson \citep{Sco2001}, and logistic \citep{ChiScott2014}. An interesting direction for future work is the development of a unified computational framework for fitting structured regression models under a wider range of distributional assumptions.

\bigskip
\begin{center}
{\large\bf SUPPLEMENTARY MATERIAL}
\end{center}

\begin{description}
\item[Title:] Supplement to ``A User-Friendly Computational Framework for Robust Structured Regression with the L$_2$ Criterion" (.tex file)
\item[Software:] \verb|L2E| R-package for performing \LTE structured regression. (GNU zipped tar file)
\end{description}

\footnotesize
\bibliographystyle{JASA}
\bibliography{ref}

\begin{thebibliography}{62}
\expandafter\ifx\csname natexlab\endcsname\relax\def\natexlab#1{#1}\fi

\bibitem[{Alfons et~al.(2013)Alfons, Croux, Gelper, et~al.}]{alfons2013sparse}
Alfons, A., Croux, C., Gelper, S. et~al. (2013), \enquote{Sparse least trimmed
  squares regression for analyzing high-dimensional large data sets,}
  \textit{The Annals of Applied Statistics}, 7, 226--248.

\bibitem[{{\'A}lvarez and Yohai(2012)}]{alvarez2012m}
{\'A}lvarez, E.~E.,  and Yohai, V.~J. (2012), \enquote{M-estimators for
  isotonic regression,} \textit{Journal of Statistical Planning and Inference},
  142, 2351--2368.

\bibitem[{Andrews(1974)}]{andrews1974robust}
Andrews, D.~F. (1974), \enquote{A robust method for multiple linear
  regression,} \textit{Technometrics}, 16, 523--531.

\bibitem[{Audibert et~al.(2011)Audibert, Catoni, et~al.}]{audibert2011robust}
Audibert, J.-Y., Catoni, O. et~al. (2011), \enquote{Robust linear least squares
  regression,} \textit{The Annals of Statistics}, 39, 2766--2794.

\bibitem[{Aybat and Wang(2016)}]{AybatWang2016}
Aybat, N.~S.,  and Wang, Z. (2016), \enquote{A Parallelizable Dual Smoothing
  Method for Large Scale Convex Regression Problems,} arXiv:1608.02227
  [math.OC].

\bibitem[{Barlow and Brunk(1972)}]{barlow1972isotonic}
Barlow, R.~E.,  and Brunk, H.~D. (1972), \enquote{The isotonic regression
  problem and its dual,} \textit{Journal of the American Statistical
  Association}, 67, 140--147.

\bibitem[{Basu et~al.(1998)Basu, Harris, Hjort, and Jones}]{BasHarHjo1998}
Basu, A., Harris, I.~R., Hjort, N.~L.,  and Jones, M.~C. (1998),
  \enquote{Robust and efficient estimation by minimising a density power
  divergence,} \textit{Biometrika}, 85, 549--559.

\bibitem[{Bertsimas and Mundru(2021)}]{Bertsimas2021}
Bertsimas, D.,  and Mundru, N. (2021), \enquote{Sparse Convex Regression,}
  \textit{INFORMS Journal on Computing}, 33, 262--279.

\bibitem[{Birke and Dette(2007)}]{birke2007estimating}
Birke, M.,  and Dette, H. (2007), \enquote{Estimating a convex function in
  nonparametric regression,} \textit{Scandinavian Journal of Statistics}, 34,
  384--404.

\bibitem[{Blanchet et~al.(2019)Blanchet, Glynn, Yan, and
  Zhou}]{blanchet2019multivariate}
Blanchet, J., Glynn, P.~W., Yan, J.,  and Zhou, Z. (2019),
  \enquote{Multivariate distributionally robust convex regression under
  absolute error loss,} in \textit{Advances in Neural Information Processing
  Systems}, pp. 11817--11826.

\bibitem[{Brunk et~al.(1972)Brunk, Barlow, Bartholomew, and
  Bremner}]{barlow1972statistical}
Brunk, H., Barlow, R.~E., Bartholomew, D.~J.,  and Bremner, J.~M. (1972),
  \enquote{Statistical inference under order restrictions: The theory and
  application of isotonic regression,} Tech. rep., Missouri Uuniversity
  Columbia Department of Statistics.

\bibitem[{Chang et~al.(2018)Chang, Roberts, and Welsh}]{chang2018robust}
Chang, L., Roberts, S.,  and Welsh, A. (2018), \enquote{Robust lasso regression
  using {T}ukey's biweight criterion,} \textit{Technometrics}, 60, 36--47.

\bibitem[{Chen and Mazumder(2021)}]{ChenMazumder2021}
Chen, W.,  and Mazumder, R. (2021), \enquote{Multivariate Convex Regression at
  Scale,} arXiv:2005.11588 [math.OC].

\bibitem[{Chi and Scott(2014)}]{ChiScott2014}
Chi, E.~C.,  and Scott, D.~W. (2014), \enquote{Robust Parametric Classification
  and Variable Selection by a Minimum Distance Criterion,} \textit{Journal of
  Computational and Graphical Statistics}, 23, 111--128.

\bibitem[{Combettes and Wajs(2005)}]{Combettes2005}
Combettes, P.~L.,  and Wajs, V.~R. (2005), \enquote{Signal Recovery by Proximal
  Forward-Backward Splitting,} \textit{Multiscale Modeling \& Simulation}, 4,
  1168--1200.

\bibitem[{Davies(1993)}]{davies1993aspects}
Davies, P.~L. (1993), \enquote{Aspects of robust linear regression,}
  \textit{The Annals of statistics}, 1843--1899.

\bibitem[{Donoho et~al.(1988)Donoho, Liu, et~al.}]{donoho1988automatic}
Donoho, D.~L., Liu, R.~C. et~al. (1988), \enquote{The ``automatic" robustness
  of minimum distance functionals,} \textit{The Annals of Statistics}, 16,
  552--586.

\bibitem[{Dykstra et~al.(1982)Dykstra, Robertson,
  et~al.}]{dykstra1982algorithm}
Dykstra, R.~L., Robertson, T. et~al. (1982), \enquote{An algorithm for isotonic
  regression for two or more independent variables,} \textit{The Annals of
  Statistics}, 10, 708--716.

\bibitem[{Ghosal and Sen(2017)}]{ghosal2017univariate}
Ghosal, P.,  and Sen, B. (2017), \enquote{On univariate convex regression,}
  \textit{Sankhya A}, 79, 215--253.

\bibitem[{Guntuboyina and Sen(2015)}]{guntuboyina2015global}
Guntuboyina, A.--- (2015), \enquote{Global risk bounds and adaptation in
  univariate convex regression,} \textit{Probability Theory and Related
  Fields}, 163, 379--411.

\bibitem[{Hannah and Dunson(2013)}]{hannah2013multivariate}
Hannah, L.~A.,  and Dunson, D.~B. (2013), \enquote{Multivariate convex
  regression with adaptive partitioning,} \textit{The Journal of Machine
  Learning Research}, 14, 3261--3294.

\bibitem[{Hjort(1994)}]{hjort1994}
Hjort, N.~L. (1994), \enquote{Minimum L2 and Robust Kullback–Leibler
  Estimation,} in \textit{Proceedings of the 12th Prague Conference}.

\bibitem[{Holland and Welsch(1977)}]{holland1977robust}
Holland, P.~W.,  and Welsch, R.~E. (1977), \enquote{Robust regression using
  iteratively reweighted least-squares,} \textit{Communications in
  Statistics-theory and Methods}, 6, 813--827.

\bibitem[{Lane(2012)}]{Lane2012}
Lane, J.~W. (2012), \enquote{Robust Quantile Regression Using L2E,} Ph.D.
  thesis.

\bibitem[{Lange(2010)}]{lange2010numerical}
Lange, K. (2010), \textit{Numerical analysis for statisticians}, Springer
  Science \& Business Media.

\bibitem[{Lange(2013)}]{lange2013optimization}
--- (2013), \textit{Optimization}, Springer, 2nd ed.

\bibitem[{Lange(2016)}]{Lange2016}
--- (2016), \textit{MM Optimization Algorithms}, Philadelphia, PA: Society for
  Industrial and Applied Mathematics.

\bibitem[{Lange et~al.(2014)Lange, Chi, and Zhou}]{LangeChiZhou2014}
Lange, K., Chi, E.~C.,  and Zhou, H. (2014), \enquote{A Brief Survey of Modern
  Optimization for Statisticians,} \textit{International Statistical Review},
  82, 46--70.

\bibitem[{Lee et~al.(1981)}]{lee1981quadratic}
Lee, C.-I.~C. et~al. (1981), \enquote{The quadratic loss of isotonic regression
  under normality,} \textit{The Annals of Statistics}, 9, 686--688.

\bibitem[{Lee(2010)}]{Lee2010}
Lee, J. (2010), \enquote{L2E estimation for finite mixture of regression models
  with applications and L2E with penalty and non-normal mixtures,} Ph.D.
  thesis.

\bibitem[{Lim(2018)}]{lim2018efficient}
Lim, C.~H. (2018), \enquote{An efficient pruning algorithm for robust isotonic
  regression,} in \textit{Advances in Neural Information Processing Systems},
  pp. 219--229.

\bibitem[{Lim and Glynn(2012)}]{lim2012consistency}
Lim, E.,  and Glynn, P.~W. (2012), \enquote{Consistency of multidimensional
  convex regression,} \textit{Operations Research}, 60, 196--208.

\bibitem[{Lin et~al.(2020)Lin, Sun, and Toh}]{LinSunToh2020}
Lin, M., Sun, D.,  and Toh, K.-C. (2020), \enquote{Efficient algorithms for
  multivariate shape-constrained convex regression problems,} arXiv:2002.11410
  [math.OC].

\bibitem[{Lozano et~al.(2016)Lozano, Meinshausen, and Yang}]{Lozano2016}
Lozano, A.~C., Meinshausen, N.,  and Yang, E. (2016), \enquote{{Minimum
  Distance Lasso for robust high-dimensional regression},} \textit{Electronic
  Journal of Statistics}, 10, 1296 -- 1340.

\bibitem[{Ma et~al.(2015)Ma, Qiu, Zhao, Ma, Yuille, and Tu}]{ma2015robust}
Ma, J., Qiu, W., Zhao, J., Ma, Y., Yuille, A.~L.,  and Tu, Z. (2015),
  \enquote{Robust $ L\_ $\{$2$\}$ E $ estimation of transformation for
  non-rigid registration,} \textit{IEEE Transactions on Signal Processing}, 63,
  1115--1129.

\bibitem[{Ma et~al.(2013)Ma, Zhao, Tian, Tu, and Yuille}]{ma2013robust}
Ma, J., Zhao, J., Tian, J., Tu, Z.,  and Yuille, A.~L. (2013), \enquote{Robust
  estimation of nonrigid transformation for point set registration,} in
  \textit{Proceedings of the IEEE conference on computer vision and pattern
  recognition}, pp. 2147--2154.

\bibitem[{Mair et~al.(2009)Mair, Hornik, and de~Leeuw}]{isotone}
Mair, P., Hornik, K.,  and de~Leeuw, J. (2009), \enquote{Isotone optimization
  in R: pool-adjacent-violators algorithm (PAVA) and active set methods,}
  \textit{Journal of Statistical Software}, 32, 1--24.

\bibitem[{Mazumder et~al.(2019)Mazumder, Choudhury, Iyengar, and
  Sen}]{mazumder2019computational}
Mazumder, R., Choudhury, A., Iyengar, G.,  and Sen, B. (2019), \enquote{A
  computational framework for multivariate convex regression and its variants,}
  \textit{Journal of the American Statistical Association}, 114, 318--331.

\bibitem[{Meng and Mahoney(2013)}]{meng2013low}
Meng, X.,  and Mahoney, M.~W. (2013), \enquote{Low-distortion subspace
  embeddings in input-sparsity time and applications to robust linear
  regression,} in \textit{Proceedings of the forty-fifth annual ACM symposium
  on Theory of computing}, pp. 91--100.

\bibitem[{Meyer(2003)}]{meyer2003test}
Meyer, M.~C. (2003), \enquote{A test for linear versus convex regression
  function using shape-restricted regression,} \textit{Biometrika}, 90,
  223--232.

\bibitem[{Ng and Maechler(2007)}]{cobs}
Ng, P.,  and Maechler, M. (2007), \enquote{A fast and efficient implementation
  of qualitatively constrained quantile smoothing splines,} \textit{Statistical
  Modeling}, 7, 315--328.

\bibitem[{Nguyen and Tran(2013)}]{Nguyen2013}
Nguyen, N.~H.,  and Tran, T.~D. (2013), \enquote{Robust Lasso With Missing and
  Grossly Corrupted Observations,} \textit{IEEE Transactions on Information
  Theory}, 59, 2036--2058.

\bibitem[{Parikh and Boyd(2014)}]{ParikhBoyd2014}
Parikh, N.,  and Boyd, S. (2014), \enquote{Proximal Algorithms,} \textit{Found.
  Trends Optim.}, 1, 127–239.

\bibitem[{Polson et~al.(2015)Polson, Scott, and
  Willard}]{PolsonScottWillard2015}
Polson, N.~G., Scott, J.~G.,  and Willard, B.~T. (2015), \enquote{{Proximal
  Algorithms in Statistics and Machine Learning},} \textit{Statistical
  Science}, 30, 559 -- 581.

\bibitem[{Ramos(2014)}]{Ramos2014}
Ramos, J.~J. (2014), \enquote{Robust Methods for Forecast Aggregation,} Ph.D.
  thesis.

\bibitem[{Riani et~al.(2014)Riani, Cerioli, Atkinson, and
  Perrotta}]{riani2014monitoring}
Riani, M., Cerioli, A., Atkinson, A.~C.,  and Perrotta, D. (2014),
  \enquote{Monitoring robust regression,} \textit{Electronic Journal of
  Statistics}, 8, 646--677.

\bibitem[{Scott(2006)}]{Scott2006}
Scott, A.~I. (2006), \enquote{Denoising by Wavelet Thresholding Using
  Multivariate Minimum Distance Partial Density Estimation,} Ph.D. thesis.

\bibitem[{Scott(1992)}]{Sco1992}
Scott, D.~W. (1992), \textit{Multivariate density estimation. Theory, practice
  and visualization}, John Wiley \& Sons, Inc.

\bibitem[{Scott(2001)}]{Sco2001}
--- (2001), \enquote{Parametric statistical modeling by minimum integrated
  square error,} \textit{Technometrics}, 43, 274--285.

\bibitem[{Scott(2009)}]{Sco2009}
--- (2009), \enquote{The {L2E} method,} \textit{Wiley Interdisciplinary
  Reviews: Computational Statistics}, 1, 45--51.

\bibitem[{Seijo and Sen(2011)}]{seijo2011nonparametric}
Seijo, E.,  and Sen, B. (2011), \enquote{Nonparametric least squares estimation
  of a multivariate convex regression function,} \textit{The Annals of
  Statistics}, 39, 1633--1657.

\bibitem[{She and Owen(2011)}]{SheOwen2011}
She, Y.,  and Owen, A.~B. (2011), \enquote{Outlier Detection Using Nonconvex
  Penalized Regression,} \textit{Journal of the American Statistical
  Association}, 106, 626--639.

\bibitem[{Stamey et~al.(1989)Stamey, Kabalin, McNeal, Johnstone, Freiha,
  Redwine, and Yang}]{prostate}
Stamey, T., Kabalin, J., McNeal, J., Johnstone, I., Freiha, F., Redwine, E.,
  and Yang, N. (1989), \enquote{Prostate specific antigen in the diagnosis and
  treatment of adenocarcinoma of the prostate II. Radical prostatectomy treated
  patients,} \textit{Journal of Urology}, 16, 1076--1083.

\bibitem[{Terrell(1990)}]{Terrell1990}
Terrell, G.~R. (1990), \enquote{Linear Density Estimates,} in
  \textit{Proceedings of the Statistical Computing Section}, American
  Statistical Association, pp. 297--302.

\bibitem[{Tibshirani(1996)}]{tibshirani1996regression}
Tibshirani, R. (1996), \enquote{Regression shrinkage and selection via the
  lasso,} \textit{Journal of the Royal Statistical Society: Series B
  (Methodological)}, 58, 267--288.

\bibitem[{Wang and Ghosh(2012)}]{wang2012shape}
Wang, J.,  and Ghosh, S.~K. (2012), \enquote{Shape restricted nonparametric
  regression with Bernstein polynomials,} \textit{Computational Statistics \&
  Data Analysis}, 56, 2729--2741.

\bibitem[{Wang et~al.(2013)Wang, Jiang, Huang, and Zhang}]{WanJiaHua2013}
Wang, X., Jiang, Y., Huang, M.,  and Zhang, H. (2013), \enquote{Robust variable
  selection with exponential squared loss,} \textit{Journal of the American
  Statistical Association}, 108, 632--643.

\bibitem[{Warwick and Jones(2005)}]{warwick2005choosing}
Warwick, J.,  and Jones, M. (2005), \enquote{Choosing a robustness tuning
  parameter,} \textit{Journal of Statistical Computation and Simulation}, 75,
  581--588.

\bibitem[{Yang et~al.(2018)Yang, Lozano, Aravkin, et~al.}]{yang2018general}
Yang, E., Lozano, A.~C., Aravkin, A. et~al. (2018), \enquote{A general family
  of trimmed estimators for robust high-dimensional data analysis,}
  \textit{Electronic Journal of Statistics}, 12, 3519--3553.

\bibitem[{Yang and Scott(2013)}]{Yang2013}
Yang, J.,  and Scott, D.~W. (2013), \enquote{Robust fitting of a Weibull model
  with optional censoring,} \textit{Computational Statistics \& Data Analysis},
  67, 149--161.

\bibitem[{Yang et~al.(2017)Yang, Pan, Yang, Zhang, Ong, and
  Tang}]{yang2017remote}
Yang, K., Pan, A., Yang, Y., Zhang, S., Ong, S.~H.,  and Tang, H. (2017),
  \enquote{Remote sensing image registration using multiple image features,}
  \textit{Remote Sensing}, 9, 581.

\bibitem[{Zou and Hastie(2005)}]{Zou2005}
Zou, H.,  and Hastie, T. (2005), \enquote{Regularization and variable selection
  via the elastic net,} \textit{Journal of the Royal Statistical Society:
  Series B (Statistical Methodology)}, 67, 301--320.

\end{thebibliography}
\end{document}